\pdfoutput=1
\RequirePackage{fix-cm}
\documentclass[reqno]{amsproc}
\usepackage{amssymb}
\usepackage{hyperref}
\usepackage{microtype}
\usepackage{lmodern}

\usepackage{tikz}
\usetikzlibrary{decorations.pathreplacing}

\usepackage{graphicx}

\usepackage[citation-order,nobysame]{amsrefs}
\usepackage{xyzbib}

\usepackage{mathtools}
\mathtoolsset{showonlyrefs}
\let\cite=\cites

\numberwithin{equation}{section}

\newcommand{\rme}{\mathrm{e}}
\newcommand{\rmi}{\mathrm{i}}
\newcommand{\rmd}{\mathrm{d}}

\newcommand{\xc}{x_\mathrm{c}}
\newcommand{\xtc}{\tilde{x}_\mathrm{c}}
\newcommand{\tc}{t_\mathrm{c}}

\newcommand{\Fhyper}[3]{ 
	\,{}_{2}F_{1}
	\left( 
		\left. \genfrac{}{}{0pt}{}{ #1 }{ #2 } \right| #3
	\right)
}

\begin{document}

\title{The five-vertex model as a discrete log-gas}

\thanks{ArXiv version of the paper published in:  
\emph{Questions of Quantum Field Theory and Statistical Physics.~31} 
(C. L. Malyshev and A. G. Pronko, eds.),
Zapiski Nauchnykh Seminarov POMI, Vol.~548, 2025, pp.~153--191 
\href{https://www.pdmi.ras.ru/znsl/2025/v548/abs153.html}{\path{https://www.pdmi.ras.ru/znsl/2025/v548/abs153.html}}. 
To be reprinted in Journal of Mathematical Sciences (eISSN
1573-8795, ISSN 1072-3374).}

\author{Filippo Colomo}
\address{INFN, Sezione di Firenze\\
Via G. Sansone 1, 50019 Sesto Fiorentino (FI), Italy}
\email{colomo@fi.infn.it}

\author{Michelangelo Mannatzu}
\address{Dipartimento di Fisica e Astronomia, Universit\`a di Firenze\\
  Via G. Sansone 1, 50019 Sesto Fiorentino (FI), Italy}
\email{michelangelo.mannatzu@gmail.com}

\author{Andrei G. Pronko} \address{St.~Petersburg Department of
  V.~A.~Steklov Mathematical Institute of the Russian Academy of
  Sciences, Fontanka 27, 191023 St.~Petersburg, Russia}
\email{a.g.pronko@gmail.com}

\begin{abstract}
We consider the five-vertex model on a rectangular domain of the
square lattice, with the so-called `scalar-product' boundary
conditions. We address the evaluation of the free-energy density of
the model in the scaling limit, that is when the number of sites is
sent to infinity and the mesh of the lattice to zero, while keeping
the size of the domain constant. To this aim, we reformulate the
partition function of the model in terms of a discrete log-gas, and
study its behaviour in the thermodynamic limit. We reproduce
previous results, obtained by using a differential equation
approach. Moreover, we provide the explicit form of the resolvent in
all possible regimes. This work is preliminary to further studies of
limit shape phenomena in the model.
\end{abstract}

\dedicatory{Dedicated to Nikolai Mikhailovich  Bogoliubov\\ 
on the occasion of his 75th birthday}


\maketitle

\section{Introduction}

The five-vertex model is a particular case of the six-vertex model,
with one of its Boltzmann weights set equal to zero
\cite{B-82,GLT-90}.  Historically, it was first introduced as a model
of crystal growth and melting, basing on the terrace-ledge-kink
picture of a crystal surface \cite{G-90}. Its exact solution and
phase diagram, in the case of periodic boundary conditions, were
worked out in \cite{GBL-93}. The model was further studied in connection
with interacting domain walls \cite{NK-94}, and as an interacting
generalization of dimer coverings of the honeycomb lattice
\cite{HWKK-96}.  Recent interest in the five-vertex model is also
based, among others, on its relation with symmetric functions
\cite{MS-13,MS-14,BBBG-19,M-20}.

Our main motivation for studying the five-vertex model stems from the
fact that, under specific choices of fixed boundary conditions, and  just
as its ascendent, the six-vertex model, it exhibits phase separation and
limit shape phenomena. In recent times, relevant progresses have been
achieved towards an understanding of the scaling properties of the
five-vertex model in a rather general setup, by variational methods
\cite{dGKW-21,KP-22a,KP-22b,KP-24}.

Here we are interested specifically in the case of `scalar-product',
or `boxed-plane-partition', boundary conditions
\cite{KBI-93,B-08,B-10,BM-15,BP-23,BM-24},
where explicit determinantal expressions of
Hankel type are known for the partition function \cite{P-16,BP-21}, and for
the boundary correlation function \cite{M-24}. These Hankel
determinant representations correspond to analogous ones obtained for
the six-vertex model with domain wall boundary conditions
\cite{K-82,I-87,BPZ-02}.  They turn out to be especially convenient,
in view of the various available techniques to study their asymptotic
behaviour in the scaling limit. In particular, explicit knowledge of
the asymptotic behaviour of the boundary correlation function would
allow for the derivation, via the `tangent method' \cite{CS-16}, of
the expression of the phase separation, or arctic curve, of the model.

In the case of the partition function of the model, its asymptotic
expansion for large lattice sizes has been worked out in
\cite{BP-24}. The derivation is based on the fact that one of the
determinant representations appears to be the $\tau$-function of the 
Painlev\'e VI. It uses an approach originally proposed in
\cite{KP-16}, based on the corresponding $\sigma$-form 
of the Painlev\'e VI 
\cite{JM-81,O-87}. The method is very efficient and systematic,
however it is unclear how it can be applied to the boundary
correlation functions.

An alternative method, inspired by \cite{Zj-00}, where it was applied
to the case of the six-vertex model, is based on random matrix model
techniques. Although less powerful 
in evaluating subleading corrections, this method appears to be more
suitable for direct extension to `modified' Hankel determinants such
as those appearing in expressions for 
boundary correlation function (see, e.g.,
\cite{CPZj-10}, for the case of the domain-wall six-vertex model).

Our goal here is therefore to work out the exact expression, in the
scaling limit, of the free-energy density of the five-vertex model
with scalar-product boundary conditions, basing on a discrete log-gas
description, and on random matrix model techniques \cite{F-10}. 
We reproduce, at the leading order, the
expressions first derived in \cite{BP-24}. On the other hand, we
evaluate the explicit form of the resolvent in the various regimes of
the considered log-gas. This is a propaedeutic result to the evaluation of the
asymptotic behaviour of the boundary correlation function, and, via
the `tangent method', of the arctic curve of the model. These two
problems will be addressed elsewhere.

We organize the paper as follows. In the next Section we review the
five-vertex model with scalar-product boundary conditions and recall
various relevant results.  
In Section 3 we reformulate the partition function of the model in terms of a
discrete log-gas, and show how to use the resolvent technique for the
evaluation of the free-energy density. The two cases where the
asymptotic form of the domain is a square or a rectangle must be
considered separately. They are treated in detail in Section 4 and 5,
respectively. Results and perspectives are discussed in Section 6.
Some technical aspects of the derivation are given in three
Appendices.


\section{The five-vertex model}

\subsection{The model}
We consider the five-vertex model on an $L\times M$ lattice
(i.e., built from the intersection of $L$ vertical and $M$ horizontal
lines), with a special choice of fixed boundary conditions, so-called
`scalar-product', or `boxed-plane-partition', boundary conditions
\cite{KBI-93,B-08,B-10,BM-15,BP-23,BM-24}.

\begin{figure}[t]
\includegraphics{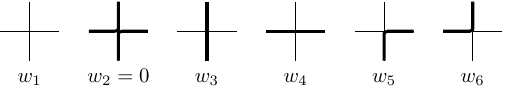}
\caption{The six vertex configurations of the six-vertex model,
  represented in terms of lines, and their Boltzmann weights in the
  five-vertex model.}\label{fig-vertices}
\end{figure}

The five-vertex model is a particular case of the six-vertex model,
with one of its Boltzmann weights set equal to zero. The
configurations of the model can be represented in terms of arrows, or
equivalently, in terms of thick or thin (or empty) edges, with the
thick edges describing paths flowing through the lattice. We will use
this last representation, with the conventions of
Fig.~\ref{fig-vertices}.

The Boltzmann weights may be parameterized as follows:
\begin{equation}
  w_1=\frac{\alpha}{\Delta}\frac{x-1}{\sqrt{x}},\qquad
  w_3=\frac{\sqrt{x}}{\alpha},\qquad w_4=\alpha \sqrt{x}, \qquad
  w_5=w_6=1.
\end{equation}
The parameter $\alpha\geq 0$ may be interpreted in terms of an
external vertical electric field. In the quantum inverse scattering 
method formalism \cite{KBI-93,BP-21} 
the parameters $x>0$ and $\Delta \in\mathbb{R}$ 
may be viewed as (the square of) a spectral
parameter and as a crossing parameter, respectively. We have
$x\in(0,1)$ for $\Delta<0$, and $x\in(1,\infty)$ for $\Delta>0$. The
parameter $\Delta$ may be defined independently of the
parametrization, as follows:
\begin{equation}
\Delta=\frac{w_3w_4-w_5w_6}{w_1w_3}.
\end{equation}
The free-fermion point of the model may be obtained, e.g., by letting
$x=\rme^{\Delta v}$ and sending $\Delta\to 0$.

\begin{figure}[t]
\includegraphics{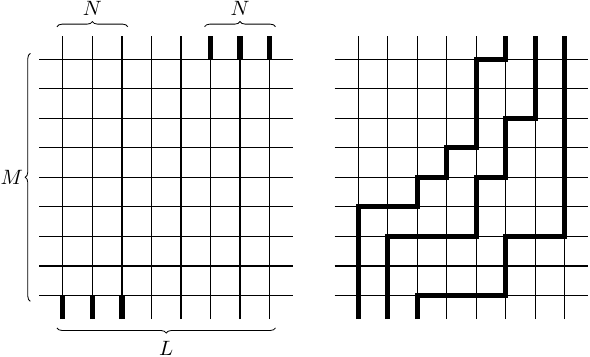}
\caption{The $L\times M$ lattice, with scalar-product boundary
  condition (left), and a possible configuration of the model
  (right). Here $L=8$, $M=9$, and $N=3$.}\label{fig-ScalarProductBC}
\end{figure}

The scalar-product boundary conditions correspond to fixing $N$ paths
entering vertically from the $N$ rightmost top boundary edges, and
exiting vertically from the $N$ leftmost bottom boundary edges, see
Fig.~\ref{fig-ScalarProductBC}.  Interestingly, the configurations of
the five-vertex model with scalar-product boundary conditions, with
$N$ lines on the $L\times M$ lattice, are in bijection with the boxed
plane partitions fitting in an $(L-N)\times N\times(M-N)$ box
\cite{B-10,P-16}. In the correspondence, the paths of the five-vertex
model are simply the `gradient lines' of the surface of the boxed
plane partition.

The partition function $Z=Z_{N,M,L}(x;\Delta,\alpha)$, is defined in
the standard way,
\begin{equation}
Z=\sum_{\mathcal{C}} \prod_{i=1,3,4,5,6}w_i^{n_i(\mathcal{C})},
\end{equation}
where the sum is taken over all admissible configurations
$\mathcal{C}$, and $n_i(\mathcal{C})$ is the number of vertices of
type $i$ in configuration $\mathcal{C}$.

Note that when the model is considered with the scalar-product 
boundary conditions, each admissible configuration $\mathcal{C}$
has the following properties:
\begin{itemize}
\item $n_1(\mathcal{C})=(L-N)(M-N)$;
  \item vertices of type 3 and 4 always appear in pairs, with
    $n_3(\mathcal{C})-n_4(\mathcal{C})=N(N+M-L)$;
\item $n_5(\mathcal{C})=n_6(\mathcal{C})$.
\end{itemize}
It follows from these conditions that the Boltzmann weight of any given
configuration $\mathcal{C}$ will be of the form
\begin{equation}
  w(\mathcal{C})=\left(\frac{\alpha}{\Delta}
  \frac{x-1}{\sqrt{x}}\right)^{(L-N)(M-N)}
  \alpha^{-N(M+N-L)}x^{[n_3(\mathcal{C})+n_4(\mathcal{C})]/2} .
\end{equation}
There are exactly $\binom{M}{N}$ configurations with maximal number of
vertices of type 3 and 4, and correspondingly, minimal number of
vertices of type 5 and 6 (with $n_5=n_6=N$). All these configurations
have Boltzmann weight
\begin{equation}
E_{N,M,L}(x;\Delta,\alpha)=\left(\frac{x-1}{\Delta}\right)^{(L-N)(M-N)}
\left(\frac{\alpha}{\sqrt{x}}\right)^{M(L-2N)}x^{N(L-N-1)}.
\end{equation}
We may thus write the partition function as
\begin{equation}\label{eq:factorZ}
Z= \binom{M}{N} E_{N,M,L}(x;\Delta,\alpha) P_{N,M,L}(x^{-1}),
\end{equation}
where $P_{N,M,L}(x^{-1})$ is a polynomial in $x^{-1}$, satisfying the
normalization condition
\begin{equation}\label{eq:Pof0}
P_{N,M,L}(0)=1.
\end{equation}
The polynomial $P_{N,M,L}(x^{-1})$ is of degree equal to the
difference between the maximal and minimal number of pairs of vertices
of type 5 and 6:
\begin{equation}\label{eq:degP}
\deg P_{N,M,L}(x^{-1})=N\min{(M-N,L-N-1)}.
\end{equation}
The polynomial $P_{N,M,L}(x^{-1})$ is symmetric under the
interchange $L \leftrightarrow M+1$,
\begin{equation}
P_{N,M,L}(x^{-1})=P_{N,L-1,M+1}(x^{-1}),
\end{equation}
however the partition function $Z$ is not.  We have no simple
explanation for this symmetry at the moment, but it is indeed evident
from the Hankel determinant representation given below.

\subsection{Particular cases}\label{sec:particular}
We report here on some particular cases which may be evaluated
explicitly in closed form, see \cite{BP-24}. These results will turn
out useful below.

Let us first consider the model at its free-fermion point, by letting
$x=\rme^{\Delta v}$, and sending $\Delta\to 0$. When the parameters
$v$ and $\alpha$ are furthermore set to 1, all weights become equals
to 1. Recalling that the configurations of the five-vertex model are
in bijection with boxed plane partitions \cite{B-10,P-16}, we simply
have
\begin{equation}
\lim_{\Delta\to 0} Z_{N,M,L}(\rme^\Delta;\Delta,1)=
\mathrm{PL}(L-N,N,M-N),
\end{equation}
where
\begin{equation}\label{eq:macmahon}
\mathrm{PL}(a,b,c)=\prod_{j=1}^a\frac{(b+c+j-1)!(j-1)!}{(b+j-1)!(c+j-1)!}
\end{equation}
is the celebrated MacMahon's formula for the number of boxed plane
partitions in a box of size $a\times b\times c$.  It is easy to check
that in the considered limit, $E_{N,M,L}(\rme^{\Delta v};\Delta,1)\to
1$, and thus, from \eqref{eq:factorZ},
\begin{equation}\label{eq:Pof1}
  P_{N,M,L}(1)=\binom{M}{N}^{-1} \mathrm{PL}(L-N,N,M-N).
\end{equation}
By using MacMahon's formula \eqref{eq:macmahon}, one may check that
 $ P_{N,M,L}(1)$ is symmetric under the interchange 
$L\leftrightarrow M+1$.

Also, letting $x=\alpha^2$ and $\Delta=-1$ in the Boltzmann weights of
the five-vertex model, in the limit $\alpha\to 0$ one recovers the
four-vertex model, whose configurations are again in bijection with
boxed plane partitions (although of different sizes), provided that
$M\geq L-1$, see \cite{B-08,BCMP-23}:
\begin{equation}
\lim_{\alpha\to0}Z(\alpha^2;-1,\alpha)=\mathrm{PL}(N,M-L+1,L-N),\qquad M\geq L-1.
\end{equation}
This observation allows us to
evaluate the leading term of the polynomial $P_{N,M,L}(x^{-1})$ in the
case $L\leq M+1$ (see also \eqref{eq:degP}):
\begin{multline}\label{eq:Pofinf1}
  \lim_{x\to 0}x^{N(L-N-1)}P_{N,M,L}(x^{-1})
\\  
=\binom{M}{N}^{-1} \mathrm{PL}(N,M-L+1,L-N),\qquad
  L\leq M+1.
\end{multline}
Note that, when setting $L=M$, the RHS simplifies to 1. Indeed, when
$L=M$, there is only one configuration maximizing the number of
vertices of type 5 and 6.  According to \cite{BP-24}, in the
complementary case $L\geq M+1$ one has
\begin{multline}\label{eq:Pofinf2}
  \lim_{x\to 0}x^{N(M-N)}P_{N,M,L}(x^{-1})
\\
=\binom{L-1}{N}^{-1}
\mathrm{PL}(N,L-M-1,M-N+1),\qquad
  L\geq M+1.
\end{multline}
It may be easily checked that indeed the last two expressions are
related through the interchange $L\leftrightarrow M+1$.

\subsection{Hankel determinant representation} 

Various determinant
representations has been worked out for the partition function of the
model \cite{BP-21,BP-24}. Here we shall resort to the following  one:
\begin{equation}
  P_{N,M,L}(x^{-1})=x^{N(N-1)/2}\,\tau_{N,M,L}(x^{-1}),
\end{equation}
with
\begin{multline}\label{eq:tau_hankel}
  \tau_{N,M,L}(x^{-1})
=N!\, C_{N,M,L}
\\ \times
  \det_{1\leq i,j,\leq N}  \left[  (x\partial_x)^{i+j-2}
  \Fhyper{-L+2,-M+1}{2}{\frac{1}{x}}
  \right],
\end{multline}
and 
\begin{equation}\label{eq:defC}
  C_{N,M,L}=\prod_{j=0}^{N-1}\frac{(L-N+j-1)!(M-N+j)!}{(L-2)!(M-1)!}.
\end{equation}
This representation, of Hankel type, has been derived in \cite{BP-24},
see Eq.~(2.6) therein.

\subsection{Scaling limit and free-energy density}

We are interested in the evaluation of the free-energy density of the
model in the scaling limit, keeping its geometry fixed. To this aim,
let us introduce the two rescaled geometric parameters
$\lambda,\mu\in[1,\infty)$, as follows: $L=\lfloor\lambda N\rfloor$,
  and $M=\lfloor\mu N\rfloor$.  We are interested in the limit
  $N\to\infty$, with $\lambda$ and $\mu$ fixed.
  
The free-energy  density of the model is defined as
\begin{equation}
F:=-\lim_{N\to\infty}\frac{1}{\lambda\mu N^2}\log Z_{N,\lfloor\mu
  N\rfloor,\lfloor\lambda
  N\rfloor}.
\end{equation}
It is given by
\begin{equation}
F=
-\frac{f_2(x)}{\mu\lambda}-\frac{(\mu-1)(\lambda-1)}{\mu
  \lambda}\log\frac{x-1}{\Delta}+\frac{\mu\lambda-2}{2\mu\lambda}\log
x-\frac{\lambda-2}{\lambda}\log\alpha,
\end{equation}
where the function 
\begin{equation}\label{eq:deff2}
f_2(x):=\lim_{N\to\infty}\frac{1}{N^2}
\log P_{N,\lfloor\mu N\rfloor,\lfloor\lambda N\rfloor}(x^{-1})
\end{equation}
describes the leading behaviour of the nontrivial factor
$P_{N,M,L}(x^{-1})$ in the partition function, see \eqref{eq:factorZ}.

As a consequence of the normalization condition \eqref{eq:Pof0}
the quantity $f_2(x)$ satisfies
\begin{equation}\label{eq:f2inf}
  \lim_{x\to\infty}f_2(x)=0.
\end{equation}
Relation \eqref{eq:Pof1} implies
\begin{equation}\label{eq:f2of1}
   f_2(1)=\Psi(\lambda-1,\mu-1),
\end{equation}
where the function $\Psi(a,b)$ is given by 
\begin{multline}\label{eq:psiab}
\Psi(a,b)  =\frac{1}{4}
\Big[\ell\left(a^2\right)-\ell\left((a+1)^2\right)
+\ell\left(b^2\right)-
    \ell\left((b+1)^2\right)
\\
-\ell\left((a+b)^2\right)+\ell\left((a+b+1)^2\right)\Big],
\end{multline}
and $\ell(x)=x\log x$, see Appendix \ref{app:psi} for details.

As a consequence of relations \eqref{eq:Pofinf1} and
\eqref{eq:Pofinf2}, we also have
\begin{equation}\label{eq:f2to0a}
\begin{split}
\lim_{x\to 0}  \left[f_2(x)+(\lambda-1)\log
    x\right]&=\Psi(\mu-\lambda,\lambda-1),\qquad \mu>\lambda,
  \\ 
  \lim_{x\to 0}
  \left[f_2(x)+(\mu-1)\log x\right]&=\Psi(\lambda-\mu,\mu-1),\qquad
  \mu<\lambda.
\end{split}
\end{equation}
Note that in the case of an asymptotically square-shaped domain,
$\lambda=\mu$, both conditions boil down to
\begin{equation}\label{eq:f2to0c}
 \lim_{x\to 0} [f_2(x)+(\lambda-1)\log
  x]=\Psi(0,\lambda-1)=0.
\end{equation}
In particular, note that the two limits, $x\to 0$ and $N\to \infty$,
do commute.

\subsection{The function $\Phi(x)$}\label{sec:Phi}
For a reason that will become apparent below, it is convenient to define
the quantity
\begin{equation}\label{eq:defPhi}
  \Phi(x):=-\lim_{N\to\infty}
  \frac{1}{N^2}\log\tau_{N,\lfloor\mu N\rfloor,\lfloor\lambda N\rfloor}(x^{-1}).
\end{equation}
A simple calculation shows that
\begin{equation}\label{eq:f2Phi}
  f_2(x):= \log\sqrt{x}-\Phi(x).
\end{equation}
Also, the behaviour of the function $\Phi(x)$ as $x\to 0, 1, \infty$
immediately follows from relations \eqref{eq:f2inf}--\eqref{eq:f2of1}
and \eqref{eq:f2to0a}--\eqref{eq:f2to0c}. One has
\begin{equation}\label{eq:phiof1}
  \Phi(1)=-\Psi(\lambda-1,\mu-1),
\end{equation}
and 
\begin{equation}\label{eq:phiinf}
\Phi(x)=\frac{1}{2}\log x+o(x^0),\qquad x\to\infty.
\end{equation}  
One also has
\begin{equation}\label{eq:phito0a}
\begin{split}
\lim_{x\to 0} \left[\Phi(x)-\left(\lambda-\frac{1}{2}\right)\log
    x\right]&=-\Psi(\mu-\lambda,\lambda-1),\qquad \mu>\lambda, 
\\ 
  \lim_{x\to 0} \left[\Phi(x)-\left(\mu-\frac{1}{2}\right)\log
    x\right]&=-\Psi(\lambda-\mu,\mu-1),\qquad
  \mu<\lambda.
  \end{split}
\end{equation}
Note that in the case of an asymptotically
square-shaped domain, $\lambda=\mu$, both conditions boil down to
\begin{equation}\label{eq:phito0c}
 \lim_{x\to 0} \left[\Phi(x)-\left(\lambda-\frac{1}{2}\right)\log
 x\right]=\Psi(0,\lambda-1)=0.
\end{equation}
In particular, note that the two limits, $x\to 0$ and $N\to \infty$,
do commute.  In the following we shall focus on the evaluation of
$\Phi(x)$.


\section{A discrete log-gas description}

\subsection{The log-gas}
We are interested in evaluating the function $f_2(x)$, or
equivalently, the function $\Phi(x)$, describing the leading behaviour
of $P_{N,M,L}(x)$ in the scaling limit. To this aim, inspired by
\cite{Zj-00}, we rewrite the Hankel determinant representation
\eqref{eq:tau_hankel} as follows:
\begin{equation}\label{eq:tau_log-gas}
  \tau_{N,M,L}(x^{-1})
  =C_{N,M,L} \sum_{0\leq k_1,\dots,k_N\leq m}\,\prod_{1\leq
    i<j\leq N}(k_j-k_i)^2
  \prod_{i=1}^N \nu_{L,M}(k_i)
\end{equation}
with $C_{N,M,L}$  defined in \eqref{eq:defC}, and 
\begin{equation}\label{eq:measure}
  \nu_{L,M}(k):=\binom{L-2}{k}\binom{M-1}{k}\frac{x^{-k} }{k+1},\qquad
  k =0,1,\dots,m,
\end{equation}
where $m:=\mathrm{min}(L-2,M-1)$.  In this formula one can easily
recognize the discrete measure analogue of an Hermitean $m\times m$
random matrix model, namely, a discrete log-gas. To investigate its
behaviour in the scaling limit, we may use standard random matrix
techniques, suitably adapted to take into account the discreteness of
the measure, see, e.g., \cite{F-10,BKMM-07} and references therein.

\subsection{Scaling limit}
We want to study the free-energy density of the considered log-gas in
the scaling limit of large $L$, $M$, $N$, with fixed ratios. To this
aim we rescale all parameters by $N$, letting $k=\lfloor Nz\rfloor$,
$M-1=\lfloor N\mu\rfloor$, $L-2= \lfloor N\lambda\rfloor$, $m= \lfloor
N\gamma\rfloor$. Here $\lambda,\mu>1$, and $\gamma=\min(\lambda,\mu)$.
Our model describes now a gas of particles with logarithmic
interaction, confined by the continuous potential
\begin{equation}
  V(z):=-\frac{1}{N}\lim_{N\to\infty}\log\nu_{\lfloor\mu N\rfloor,
    \lfloor\lambda N\rfloor}(Nz).
\end{equation}
In our case, it is given by
\begin{multline}\label{eq:potential}
V(z) =2z\log
  z+(\lambda-z)\log(\lambda-z)-\lambda\log \lambda
\\ 
+(\mu-z)\log(\mu-z)-\mu\log \mu+z\log x,   
\end{multline}
complemented by two hard walls, at $z=0$, and at
$z=\gamma$.

To proceed, we rely on standard results from weighted potential theory
in the case of a discrete log-gas \cite{BKMM-07}. The problem may be
reformulated as the evaluation of the equilibrium measure $\rho(z)$,
which is the unique minimizer of the functional
\begin{equation}
  S[\rho]=-\int\int_{z\not=w}\log|w-z|\rho(w)\rho(z)\rmd w\rmd z+\int V(z)
  \rho(z)\rmd z,
\end{equation}
subject to the normalization condition
\begin{equation}\label{eq:normalization}
  \int\rho(z)\rmd z=1
\end{equation}
and the additional constraint 
\begin{equation}\label{eq:constraint}
  0\leq \rho(z)\leq 1,
\end{equation}
that follows from the discreteness of the original variables $k_j$
\cite{DK-93}. In the above formulae, integration is understood over
the interval $[0,\gamma]$.

As a consequence of the last constraint, the equilibrium measure
partitions the interval $[0, \gamma]$ into a sequence of intervals,
known as \emph{voids}, \emph{bands}, and \emph{saturated regions},
according to its value: $\rho(z)=0$, or $0<\rho(z)<1$, or $\rho(z)=1$,
respectively. Voids and saturated intervals go together under the
name of \emph{gaps}.

\subsection{The resolvent}
A convenient way to evaluate the equilibrium measure is based on the
resolvent,
\begin{equation}
  W(z)=\int_S\frac{\rho(u)}{z-u}\rmd u, \qquad z\not\in S
\end{equation}
where $S$ is the support of the equilibrium measure $\rho(z)$. The
resolvent has the following properties:
\begin{itemize}
\item[\emph{i)}] it is analytic in $\mathbb{C}\setminus S$;
\item[\emph{ii)}] the normalization condition \eqref{eq:normalization}
  implies the asymptotic behaviour:
  \begin{equation}\label{eq:largez}
    W(z)\sim\frac{1}{z}, \qquad z\to\infty;
  \end{equation} 
\item[\emph{iii)}] the equilibrium measure and the resolvent are
  related by
 \begin{equation}
  \rho(z)=-\frac{1}{2\pi\rmi}\left[W(z+\rmi 0)-W(z-\rmi 0)\right],
  \qquad z\in S;
 \end{equation} 
\item[\emph{iv)}] the resolvent satisfies the `saddle-point equation':
   \begin{equation}\label{eq:WSPE}
     W(z+\rmi 0)+W(z-\rmi 0)=U(z), \qquad z\in S.
     \end{equation} 
  \end{itemize}
In the case of a continuous log-gas, $U(z)$ is just equal to the
derivative $V'(z)$ of the potential of the model. However, in the case
of a discrete log-gas, this holds only as long as the equilibrium
measure $\rho(z)$ does not saturate the constraint
\eqref{eq:constraint}. The occurrence of saturation in the equilibrium
measure requires modifying the form of $U(z)$ and $S$ in
\eqref{eq:WSPE}, as we shall discuss later on.

Assuming that the support $S$ consists in one single interval $[a,b]$
on the real axis, the solution of \eqref{eq:WSPE} is
\begin{multline}\label{eq:Wsol}
  W(z)=\frac{\sqrt{(z-a)(z-b)}}{2\pi}\int_a^b
  \frac{U(u)}{(z-u)\sqrt{(u-a)(b-u)}}\rmd u,
  \\ 
  z\in\mathbb{C}\setminus [a,b].
  \end{multline}
Imposing that the expansion at large $z$ of the explicit expression of
the resolvent, obtained by evaluating the integral in \eqref{eq:Wsol},
matches the asymptotic behaviour \eqref{eq:largez} provides two
conditions, that determine the end-points $a$ and $b$ of the support
$S$.

\subsection{The free-energy density}
The free-energy density of our discrete log-gas coincides with the
function $\Phi(x)$ defined in \eqref{eq:defPhi}.

Denoting by $E$ the first moment of the equilibrium measure $\rho(u)$,
or equivalently, the average position of the log-gas particles, we
note that it can be worked out from the asymptotic expansion of the
resolvent. Indeed, one has
\begin{equation}
  W(z)=\frac{1}{z}+\frac{E}{z^2}+O(z^{-3}),\qquad |z|\to\infty.
\end{equation}
We also note that, due to the form of the potential
\eqref{eq:potential}, the free-energy density $\Phi(x)$ may be related
to the first moment $E$ by
\begin{equation}\label{eq:EPhi}
  x\partial_x\Phi(x) = E.
\end{equation}
The log-gas free-energy density $\Phi(x)$ may thus be evaluated up to
an integration constant, which may be fixed if one is able to
calculate explicitly $\Phi(x)$ at some particular value of $x$ by
other means. Such calculation has indeed been done in Section
\ref{sec:Phi}, yielding conditions
\eqref{eq:phiof1}-\eqref{eq:phito0c}, at $x=1$, or as $x\to 0,\infty$.


\section{The case $\lambda=\mu$}

As already emphasized in \cite{BP-24}, the free-energy density of the
five-vertex model exhibits different behaviours according to the
asymptotic shape (square or rectangular) of the considered domain.
This happens of course in the free-energy density of the corresponding
log-gas as well, depending on the values of $\lambda$ and $\mu$ being
equal or different. The two cases must be treated separately; we start
with the simplest one, where $\lambda=\mu$.

\subsection{Preliminaries}
When $L=M-1$, the measure \eqref{eq:measure} simplifies
significantly. In the scaling limit, such simplification occurs as
soon as $L-M=o(N)$. In this case, the corresponding potential is
similar to that associated to the orthogonalizing measure for
Krawtchouk orthogonal polynomials. The asymptotic behaviour of the
Krawtchouk log-gas has been studied in \cite{DS-00}. Note however here
an overall factor 2 in the potential. As we shall see below, this
yields significant differences in the asymptotic behaviour of the
log-gas, with respect to the Krawtchouk case, in particular, the
emergence of two third-order phase transitions in the free-energy
density of the model.

In the scaling limit we have $\lambda=\mu$. The
potential becomes
\begin{equation}
  V(z)=2[z \log z
    +(\lambda-z)\log(\lambda-z)-\lambda\log\lambda]+z\log x,
  \end{equation}
with two additional hard walls located at $z=0$ and $z=\lambda$. The parameter
$x>0$ enters through its logarithm. When $x=1$, the potential is
symmetric under $z\leftrightarrow\lambda -z$, with its minimum at
$\lambda/2$. When $x\not=1$, the minimum of the potential,
$z_m=\lambda/(\sqrt{x}+1)$ is displaced to the right (if $0<x<1$) or
left (if $x>1$).

The shape of the potential, and the presence of the constraint
\eqref{eq:constraint} suggest various possible scenarios, built as
suitable sequences of saturated intervals (S), bands (B), and voids
(V). We shall denote each scenario by the initials of the sequence,
from left to right, of these possibile behaviours of the density.
Simplest scenarios have only one band.  The presence of two bands
would imply the necessity to solve a two-cut problem. Luckily, only
one-band scenarios will occur here.  More precisely, decreasing $x$
within the interval $(0,\infty)$, three possible scenarios will
emerge, namely SBV, VBV, and VBS.  These three scenarios are separated
by two critical values of $x$. In correspondence of these two values
the free-energy density exhibits a discontinuity in its third
derivative, and the log-gas experiences a third-order phase
transition.

\subsection{The void-band-void scenario (VBV)}
For large enough values of $\lambda$, and small enough values of
$|\log x|$, we expect the particles to stay far away from the
hard walls, and with relatively small density. In other words, we expect
a single central band, $[a,b]$, between two voids, $[0,a]$ and
$[b,\lambda]$. We shall denote such scenario by the acronym
VBV. Calculations may be performed under the one-cut assumption.

\subsubsection{The resolvent}
We need to solve the saddle-point equation \eqref{eq:WSPE} where in
the presently considered case the RHS reads
\begin{equation}
  U(z)=V'(z)=2\log z- 2\log(\lambda-z)+\log x.
\end{equation}
Inserting into \eqref{eq:Wsol} and using the integration fomulae given in
Appendix \ref{app:integrals}, we obtain the following expression for
the resolvent:
\begin{equation}\label{eq:ressymVBV}
  W(z)=\log \sqrt{x} +2\log
  \dfrac{\sqrt{a(z-b)}+\sqrt{b(z-a)}}{\sqrt{(\lambda-a)(z-b)}+
    \sqrt{(\lambda-b)(z-a)}}.
\end{equation}
Imposing the correct asymptotic behaviour provides the two equations
\begin{align}
\frac{\sqrt{\lambda-a}+\sqrt{\lambda-b}}{\sqrt{a}+\sqrt{b}} &= x^{1/4},
  \\
 \lambda-\sqrt{ab}-\sqrt{\lambda-a}\sqrt{\lambda-b}&=1,
\end{align}
whose solution determines the end-points $a$ and $b$, given by 
\begin{equation}\label{eq:endpointsVBV}
  a=\frac{(\sqrt{2\lambda-1}-x^{1/4})^2}{2(1+\sqrt{x})},\qquad
  b=\frac{(\sqrt{2\lambda-1}+x^{1/4})^2}{2(1+\sqrt{x})}.
\end{equation}
The corresponding equilibrium measure reads
\begin{equation}\label{eq:rhoVBV}
  \rho(z)=\frac{2}{\pi}
  \arctan\sqrt{\frac{(\lambda-a)(b-z)}{(\lambda-b)(z-a)}}-
    \frac{2}{\pi}
  \arctan\sqrt{\frac{a(b-z)}{b(z-a)}},\qquad z\in[a,b],
\end{equation}
with $a$ and $b$ given by \eqref{eq:endpointsVBV}.

\subsubsection{The end-points of the band}
The present derivation and results are valid as long as the VBV
scenario holds. A first condition for this is that the equilibrium
measure \eqref{eq:rhoVBV}, with $a$ and $b$ given by
\eqref{eq:endpointsVBV}, satisfies the condition $\rho(z)<1$. It is a
simple calculation to verify that this indeed holds.

Also, the end-points \eqref{eq:endpointsVBV} must satisfy the
conditions $0<a$ and $b<\lambda$. More precisely, we expect the VBV
scenario to break down when $a=0$, that is when $x=\xc$, with
\begin{equation}\label{eq:xcrit}
\xc:=(2\lambda-1)^2,
\end{equation}
or when $b=\lambda$, which occurs when $x=\xtc$, with
$\xtc=(2\lambda-1)^{-2}\equiv\xc^{-1}$. These two critical values
separates three different regimes, labelled as follows. Regime I:
$x>\xc$, Regime II: $\xtc<x<\xc$, and Regime III: $x<\xtc$,
corresponding to the SBV, VBV, and VBS scenarios, respectively.

\subsubsection{The free-energy density}
We now turn to the evaluation of the free-energy density in the
presently considered scenario, VBV, corresponding to values
$x\in[\xtc,\xc]$.  Expanding the resolvent \eqref{eq:ressymVBV} to
order $z^{-2}$, we get
\begin{equation}
  E_\mathrm{VBV}=
  \frac{(a+b)(2\lambda-1)+\lambda(\sqrt{a}+\sqrt{b})^2}{4\sqrt{x}}
  \left(\frac{\sqrt{b}-\sqrt{a}}{\sqrt{b}+\sqrt{a}}\right)^2.
\end{equation}
Replacing herein the expressions for the end-points
\eqref{eq:endpointsVBV}, we get the first moment of the density:
\begin{equation}
E_\mathrm{VBV}=\frac{2\lambda-1}{2(1+\sqrt{x})}+\frac{1}{4},
\end{equation}
which, recalling \eqref{eq:EPhi}, implies
\begin{align}
  \Phi_\mathrm{VBV}(x)&=\frac{1}{4}\log x
  +(2\lambda-1)\log\frac{\sqrt{x}}{1+\sqrt{x}}+C
  \\&=\frac{1}{4}\log x
  +(2\lambda-1)\log\frac{2\sqrt{x}}{1+\sqrt{x}} -\Psi(\lambda-1,\lambda-1),
\end{align}
where the integration costant $C$ has been determined by the condition
$\Phi(1)=-\Psi(\lambda-1,\lambda-1)$, see
\eqref{eq:phiof1} and \eqref{eq:Psill}.
The free-energy density of the
log-gas for $x\in[\xc^{-1},\xc]$ may be equivalently   written
\begin{equation}\label{eq:PhisymVBV}
  \Phi_\mathrm{VBV}(x)=\left(\lambda-\frac{1}{4}\right)\log\frac{x}{\xc}
    -(2\lambda-1)\log\frac{1+\sqrt{x}}{1+\sqrt{\xc}} +\Phi_\mathrm{c},
\end{equation}
where
\begin{equation}\label{eq:phicrit}
  \Phi_\mathrm{c}:=\Phi_\mathrm{VBV}(\xc)=\frac{1}{2}
  \log \xc-(\lambda-1)^2\log\frac{\xc}{\xc-1},
\end{equation}
is the value of free-energy density at the critical point.

\subsection{The saturated-band-void scenario (SBV)}
We know that for values of $x$ close to 1 the particles of the log-gas
accumulate around the minimum of the potential, with a density that
never saturates the constraint $\rho(u)=1$.  As we increase the value
of $x$, the minimum of the potential moves to the left, and at some
value $\xc$ we may expect the left end-point of the support of the
equilibrium measure to touch the left hard wall, with the emergence of
a saturated interval for values $x>\xc$, and hence, the transition to
the SBV scenario.

\subsubsection{The resolvent}
The present scenario consists in a saturated interval $[0,a]$, a band
$[a,b]$, and a void $[b,\lambda]$. In correspondence of the saturated
interval $[0,a]$, where the equilibrium measure evaluates to one, the
resolvent has a logarithmic cut with discontinuity $-2\pi\rmi$. This
can be removed by introducing an auxiliary function $H(z)$ as
follows,
\begin{equation}
W(z)=\log\frac{z}{z-a}+H(z). 
\end{equation}
Since $W(z)$ must solve   \eqref{eq:WSPE}, with
$U(z)=V'(z)$, it follows that $H(z)$ must satisfy
\begin{equation}
  H(z+\rmi 0)+H(z-\rmi 0)= \log x + 2\log\frac{z-a}{\lambda-z},
  \qquad z\in[a,b].
\end{equation}
The auxiliary resolvent $H(z)$ may now be evaluated in the standard
way, using relation \eqref{eq:Wsol} and the identities of Appendix
\ref{app:integrals}. We finally get:
\begin{equation}\label{eq:ressymSBV}
  W(z)=\log\sqrt{x}+2\log\frac{\sqrt{b-a}\sqrt{z}}
  {\sqrt{\lambda-a}\sqrt{z-b}+\sqrt{\lambda-b}\sqrt{z-a}}.
\end{equation}
Requiring the asymptotic behaviour $W(z)\sim 1/z$, as $z\to\infty$, we
obtain the two equations
\begin{align}
\frac{\sqrt{\lambda-a}+\sqrt{\lambda-b}}{\sqrt{b-a}}&=x^{1/4},\\
  \lambda-\sqrt{\lambda-a}\sqrt{\lambda-b}&=1.
\end{align}
whose solution determines the end-points of the band.  For $x>1$,
which is definitely holding in the presently considered case, namely
$x>\xc$, we have
\begin{align}
a&=\frac{\sqrt{x}+1-2\lambda}{\sqrt{x}-1},
  \\
b&=\frac{\sqrt{x}-1+2\lambda}{\sqrt{x}+1}.
\end{align}
It is easily checked that the condition $a=0$ reproduces the critical
value $\xc=(2\lambda-1)^2$.

The equilibrium measure  reads
\begin{equation}
  \rho(z)=
  \frac{2}{\pi}\arctan\sqrt{\frac{(\lambda-a)(b-z)}{(\lambda-b)(z-a)}},
  \qquad z\in[a,b],
\end{equation}
with $\rho(z)=1$ for $z\in[0,a]$ and $\rho(z)=0$ for $z>b$.

\subsubsection{The free-energy density}
Expanding the resolvent \eqref{eq:ressymSBV} to order $z^{-2}$, we get
the first moment of the equilibrium measure
\begin{align}
  E_\mathrm{SBV}&=\frac{1}{4}
  \left[2\lambda(\lambda-\sqrt{\lambda-a}\sqrt{\lambda-b})
    -(a+b)\sqrt{\lambda-a}\sqrt{\lambda-b}\right]
  \\&= \frac{1}{2}+\frac{(\lambda-1)^2}{x-1}.
\end{align}
Integrating, see \eqref{eq:EPhi}, we obtain the free-energy density
\begin{equation}\label{eq:PhisymSBV}
  \Phi_\mathrm{SBV}(x)=\frac{1}{2}\log x-(\lambda-1)^2\log\frac{x}{x-1}.
\end{equation}
The condition \eqref{eq:phiinf} fixes the integration constant simply
to zero.  Note that the value $\Phi_\mathrm{SBV}(\xc)$ coincides with
$\Phi_\mathrm{c}:=\Phi_\mathrm{VBV}(\xc)$, reported in
\eqref{eq:phicrit}.  The function $\Phi(x)$ is thus continuous at
$x=\xc$, as expected.  It may be verified that the first and second
derivatives of $\Phi(x)$ are continuous as well, over the whole
interval $(\xtc,\infty)$, while the third derivative is discontinuous
at $x=\xc$. In other words, we observe the occurrence of a third-order
phase transition in the free-energy density of the log-gas at $x=\xc$.

\subsection{The void-band-saturated scenario (VBS)}
We now turn to values of the parameter $x$ in the interval
$x<\xtc$. As seen above, when $x$ tends to $\xtc$ from above, the
right end-point $b$ of the band gets closer and closer to the hard
wall located at $z=\lambda$. Decreasing the parameter $x$ to values
$x<\xtc$, a saturated region arises to the right of the band, with the
emergence of a VBS scenario.

\subsubsection{The resolvent}
In the presently considered scenario, consisting in a void $[0,a]$, a
band $[a,b]$, and a saturated interval $[b,\lambda]$, the resolvent
has a logarithmic cut in correspondence of the saturated interval,
which can be removed by introducing an auxiliary function $H(z)$ as
follows:
\begin{equation}
  W(z)=\log\frac{z-b}{z-\lambda}+H(z).
\end{equation}
This function must satisfy
\begin{equation}
  H(z+\rmi 0)+ H(z-\rmi 0)=\log x+2\log\frac{z}{b-z},
  \qquad z\in[a,b].
\end{equation}
The expression for the resolvent is therefore
\begin{equation}\label{eq:ressymVBS}
  W(z)=\log\sqrt{x}+2\log
  \frac{\sqrt{a}\sqrt{z-b}+\sqrt{b}\sqrt{z-a}}{\sqrt{b-a}\sqrt{z-\lambda}},
\end{equation}
yielding the equations for the end-points
\begin{align}
  \frac{\sqrt{b-a}}{\sqrt{a}+\sqrt{b}}&=x^{1/4}, \\
  \lambda-\sqrt{ab}&=1,
  \end{align}
whose solution is (for $x<1$, which holds in the currently considered
scenario):
\begin{align}
  a&=\frac{1-\sqrt{x}}{1+\sqrt{x}}(\lambda-1),
  \\
  b&=\frac{1+\sqrt{x}}{1-\sqrt{x}}(\lambda-1).
\end{align}
It is easily checked that the condition $b=\lambda$ reproduces the
critical value $\xtc=(2\lambda-1)^{-2}$.

The  equilibrium measure  reads
\begin{equation}
  \rho(z)=\frac{2}{\pi} \arctan\sqrt{\frac{b(z-a)}{a(b-z)}}, 
  \qquad z\in[a,b].
\end{equation}
with $\rho(z)=0$ for $z\in[0,a]$ and $\rho(z)=1$ for
$z\in[b,\lambda]$.  It is worth noting that all results worked out
here for the VBS scenario can be obtained from those for the SBV
scenario by means of the substitution $x\to x^{-1}$, $a\to\lambda-b$,
and $b\to\lambda-a$.

\subsubsection{The free-energy density}
Expanding the resolvent \eqref{eq:ressymVBS} to order $z^{-2}$, we get
\begin{align}
  E_\mathrm{VBS}&=\frac{1}{4}
  \left[2\lambda^2
    -(a+b)\sqrt{ab}\right]
  \\&= \frac{\lambda^2}{2}-\frac{1}{2}(\lambda-1)^2\frac{1+x}{1-x},
\end{align}
which in turn implies:
\begin{equation}\label{eq:PhisymVBS}
  \Phi_\mathrm{VBS}(x)=\frac{2\lambda-1}{2}\log x+(\lambda-1)^2\log(1-x).
\end{equation}
The condition \eqref{eq:phito0c} fixes the integration constant simply
to zero.  It may be verified that the value $\Phi_\mathrm{VBS}(\xtc)$
coincides with $\Phi_\mathrm{VBV}(\xtc)$, and that the function
$\Phi(x)$ is continuous at $x=\xtc$, as expected. It is just a matter
of calculation to check that the first and second derivatives of
$\Phi(x)$ are continuous as well, over the whole interval
$(0,\infty)$, while the third derivative is discontinuous at $x=\xtc$.

Summing up, we have determined the free-energy density $\Phi(x)$ of
the considered log-gas in the scaling limit, that is when $M-L=o(N)$
as $N\to\infty$. The expression of the free-energy density indicates
that the system undergoes two third-order phase transitions, at
$x=\xtc$ and $x=\xc$, with $\xc=(2\lambda-1)^2=\xtc^{-1}$.


\section{The case $\lambda\not=\mu$}

We now consider the same log-gas in a different scaling limit, namely
when $M-L=O(N)$ as $N\to \infty$. Concerning the rescaled parameters,
this implies $\lambda\not=\mu$. The situation appears to be
significantly more complicated, with respect to the previous case,
$\lambda=\mu$. As we shall see, four different scenarios appear. They
all have a single band $[a,b]$ (thus falling into the class of one-cut
problems), between two gaps $[0,a]$ and $[b,\gamma]$, where
$\gamma=\min(\lambda,\mu)$, which can be either voids or saturated
regions.

Despite the presence of four different scenarios for the equilibrium
measure of the log-gas, only two different phases, or regimes, will
appear, with a single critical point. A similar situation was observed
in \cite{CP-15}, where a transition between two different scenarios
could occur, with the free-energy density staying continuous together
with all its derivatives, and hence without any phase transition, or
change of regime.

\subsection{Preliminaries}
When $\lambda\not=\mu$, the potential of our log-gas is given by
\eqref{eq:potential}, with $z\in[0,\gamma]$, where
$\gamma=\min{(\lambda,\mu)}$, with two hard walls at $z=0$, and at
$z=\gamma$.

The derivative of the potential reads:
\begin{equation}\label{eq:derivpot}
V'(z)=2\log z-\log(\lambda-z)-\log(\mu-z)+\log x.
\end{equation}
Just as in the case $\mu=\lambda$, the derivative of the potential
diverges in correspondence of the two hard walls, and the potential
profile joins up smoothly to the hard walls on both sides. 
Letting $\lambda<\mu$ for definiteness, this is 
due to the terms `$\log z$' and `$\log(\lambda-z)$', in
\eqref{eq:derivpot}, for the left and right hard wall, respectively.
Note however that now the coefficient of the term associated to the
right hard wall is exactly $1$.  As we shall see below, this will imply,
at variance with the symmetric case discussed in the previous Section,
different behaviours in the free-energy density when, under variation
of the parameters of the potential, the right end-point of the support
of the equilibrium measure reaches the right hard wall.

The derivative of the potential vanishes at $z=z_{\pm}$, with
\begin{equation}
  z_{\pm}:=\frac{-(\lambda+\mu)
    \pm\sqrt{(\lambda+\mu)^2+4\lambda\mu(x-1)}}{2(x-1)}.
\end{equation}
When $x>1$, $z_-$ is always negative, for any $\mu,\lambda$. Vice
versa, when $x<1$, $z_-$ is positive, but also always larger than
$\gamma$. In both cases, only $z_+$ lies inside inside the interval
$[0,\gamma]$, and actually strictly inside for any
$x\in(0,\infty)$. It easy to verify that at $z=z_+$ the potential has
a minimum.

Restricting to one-band scenarios, which will turn out to be
sufficient for our purposes, the form of the potential suggests to
investigate the four cases with a central band on the interval
$[a,b]$, between two external intervals $[0,a]$, and $[b,\gamma]$,
each of which may be saturated or void. Specifically we will discuss
below the SBV, SBS, VBV, and VBS scenarios, in this order.

The potential being symmetric under interchange of the parameters
$\lambda$ and $\mu$, we hereafter assume, with no loss of generality,
that $\lambda<\mu$, and get rid of the parameter $\gamma$. 

\subsection{The saturated-band-void scenario (SBV)}
To start with, let us consider the case of relatively large $x$. The
minimum of the potential, $z_+$, is then relatively close to the
origin, and the particles tend to accumulate to the left. It is
reasonable to expect an SBV scenario, with a saturated interval
$[0,a]$, a band $[a,b]$ and a void $[b,\lambda]$. 

\subsubsection{The resolvent and the band end-points}
Along with our usual procedure, we have 
\begin{equation}
  W(z)=\log\frac{z}{z-a}+H(z),
\end{equation}
where the auxiliary resolvent $H(z)$ must now satisfy
\begin{equation}
H(z+\rmi 0)+H(z-\rmi 0)=V'(z)-2\log\frac{z}{z-a},
\end{equation}
or, explicitly,
\begin{equation}
H(z+\rmi 0)+H(z-\rmi 0)
=\log x+ 2\log(z-a)-\log(\mu-z)-\log(\lambda-z).
\end{equation}
Standard calculations yield the following expression for the  resolvent:
\begin{multline}\label{eq:resasymSBV}
  W(z)=\log\sqrt{x}
  +\log\frac{\sqrt{b-a}\sqrt{z}}{\sqrt{\lambda-a}\sqrt{z-b}+
    \sqrt{\lambda-b}\sqrt{z-a}}\\
  +\log\frac{\sqrt{b-a}\sqrt{z}}{\sqrt{\mu-a}\sqrt{z-b}
    +\sqrt{\mu-b}\sqrt{z-a}}.
\end{multline}
Requiring the asymptotic behaviour $W(z)\sim 1/z$ as $z\to\infty$, we
obtain the two equations
\begin{align}\label{eq:eqnabSBV1}
  \frac{\sqrt{\mu-a}+\sqrt{\mu-b}}
  {\sqrt{\lambda-a}-\sqrt{\lambda-b}}&=\sqrt{x},
  \\
  \mu+\lambda-\sqrt{\mu-a}\sqrt{\mu-b}-\sqrt{\lambda-a}\sqrt{\lambda-b}&=2.
  \label{eq:eqnabSBV2}
\end{align}
Recalling that $\lambda<\mu$, it is evident from the first one that
the currently considered scenario, SBV, may occur only for $x>1$.  The
solution of these two equations provides the end-points of the band,
\begin{align}\label{eq:abSBV1}
  a&=\frac{x+1-\mu-\lambda-2\sqrt{x(\mu-1)(\lambda-1)}}
  {(x - 1)},
\\
  b&=\frac{x+1-\mu-\lambda+2\sqrt{x(\mu-1)(\lambda-1)}}
  {(x - 1)}.\label{eq:abSBV2}
\end{align}
where some simplifications have been done under the assumption $x>1$,
holding in the present scenario.

Clearly, the currently considered scenario, SBV, may hold only as long
as $a\geq 0$, and $b\leq \lambda$. The first condition implies $x\geq
\xc$, where
\begin{equation}\label{eq:xcrit2}
\xc:= \left(\sqrt{\lambda\mu}+\sqrt{(\lambda-1)(\mu-1)}\right)^2.
\end{equation}
It is easily checked that $\xc>1$. As for the second condition, it
implies $x>x_1$, with
\begin{equation}\label{eq:x1}
   x_1=\frac{\mu-1}{\lambda-1}.
\end{equation}
It is easily checked that $x_1>1$, as well. Therefore the current
scenario, SBV, holds for large $x$, down to the larger value between
$x_1$ and $\xc$, that is for $x>\max(x_1,\xc)$.
    
In the SBV scenario, the equilibrium measure reads
\begin{equation}
  \rho(z)=\frac{1}{\pi}
  \arctan\sqrt{\frac{(\lambda-a)(b-z)}{(\lambda-b)(z-a)}}
  +\frac{1}{\pi} \arctan\sqrt{\frac{(\mu-a)(b-z)}{(\mu-b)(z-a)}}
\end{equation}
for $z\in[a,b]$, with $\rho(z)=1$ for $z\in[0,a]$, and $\rho(z)=0$ for
$z>b$.

\subsubsection{The free-energy density}
Expanding the resolvent \eqref{eq:resasymSBV} to order $z^{-2}$, we
get
\begin{multline}
  E_\mathrm{SBV}=\frac{1}{8}
  \Big[2\lambda(\lambda-\sqrt{\lambda-a}\sqrt{\lambda-b})
    +2\mu(\mu-\sqrt{\mu-a}\sqrt{\mu-b})
\\
 -(a+b)
    \left(\sqrt{\lambda-a}\sqrt{\lambda-b}+\sqrt{\mu-a}\sqrt{\mu-b}\right)\Big].
\end{multline}
This amounts to
\begin{equation}\label{eq:ESBV}
 E_\mathrm{SBV} 
=\frac{(\lambda-1)(\mu-1)}{x-1}+\frac{1}{2},\qquad x>\max(x_1,\xc),
\end{equation}
which in turn implies the free-energy density
\begin{equation}\label{eq:PhiSBV}
  \Phi_\mathrm{SBV}(x)= \frac{1}{2}\log x-(\lambda-1)(\mu-1)\log\frac{x}{x-1}.
  \end{equation}
The condition \eqref{eq:phiinf} fixes the integration constant simply
to zero. 

\subsection{The saturated-band-saturated  scenario (SBS)}
When $\lambda$ is relatively close to 1, and $x$ tuned to have the
minimum of the potential centered around $\lambda/2$, the SBS scenario
may in principle occur. Let us investigate this possibility, with a
band $[a,b]$ and two saturated intervals, $[0,a]$ and $[b,\lambda]$.

\subsubsection{The resolvent} Following our standard procedure,
we introduce an auxiliary resolvent $H(z)$, defined by
\begin{equation}
W(z)=\log\frac{z}{z-a}+\log\frac{z-b}{z-\lambda}+H(z),
\end{equation}
and determined by the `potential'
\begin{equation}
V'(z)-2\log\frac{z}{z-a}-2\log\frac{b-z}{\lambda-z}, \qquad z\in[a,b].
\end{equation}
Using the relations reported in Appendix \ref{app:integrals} we obtain
the following expression for the resolvent:
\begin{multline}\label{eq:resasymSBS}
  W(z)=\log\sqrt{x}+\log\frac{\sqrt{b-a}\sqrt{z}}
  {\sqrt{\lambda-a}\sqrt{z-b}-\sqrt{\lambda-b}\sqrt{z-a}}
  \\
  +\log\frac{\sqrt{b-a}\sqrt{z}}
           {\sqrt{\mu-a}\sqrt{z-b}+\sqrt{\mu-b}\sqrt{z-a}}.
\end{multline}
Note that this expression for the resolvent may be formally obtained
from that of the SBV scenario, Eq. \eqref{eq:resasymSBV}, by means of
the replacement $\sqrt{\lambda-b}\to - \sqrt{\lambda-b}$.  Requiring
the asymptotic behaviour $W(z)\sim 1/z$ as $z\to\infty$, we obtain the
two equations
\begin{align}
\frac{\sqrt{\mu-a}+\sqrt{\mu-b}}{\sqrt{\lambda-a}+\sqrt{\lambda-b}}
  &=\sqrt{x}, 
\\ 
\mu+\lambda-\sqrt{\mu-a}\sqrt{\mu-b}+\sqrt{\lambda-a}\sqrt{\lambda-b}&=2,
\end{align}
whose solution determines the position of the end-points $a$ and
$b$. Again note that these can be obtained from the corresponding
equations \eqref{eq:eqnabSBV1} and \eqref{eq:eqnabSBV2} of the SBV
scenario, by means of the replacement $\sqrt{\lambda-b}\to -
\sqrt{\lambda-b}$.  Note also that, since $\lambda<\mu$, it follows
from the first equation that the present scenario may occur only for
values of $x>1$. Furthermore, comparison with \eqref{eq:eqnabSBV1}
shows that it may occur for smaller values of $x$, with respect to the
SBV scenario.

It appears that the solutions of these equations are given by
\eqref{eq:abSBV1} and \eqref{eq:abSBV2}, just as in the SBV scenario,
provided that $x<x_1$.  Clearly, decreasing the value of $x$, the SBV
scenario will hold until $x$ reaches the larger value between $x_1$
and $\xc$. If $x_1>\xc$, the transition occurs at $x=x_1$, to the SBS
scenario. If, instead, $x_1<\xc$, the transition occurs at $x=\xc$,
towards the VBV scenario.  In other words, the SBS scenario may occur
only if $x_1>\xc$. In that event, it actually holds for all
$x\in[\xc,x_1]$.

The requirement  $x_1>\xc$ selects the region 
\begin{equation}\label{eq:x1gtxc}
  1< \lambda <\frac{4}{3}, \qquad
    \mu>\frac{(2-\lambda)^2}{4-3\lambda},
  \end{equation}
in the space of parameters $\mu>\lambda>1$. In conclusion, the SBS
scenario occurs if and only if the parameters $\lambda$ and $\mu$
satisfy the condition \eqref{eq:x1gtxc}. In this case the equilibrium
measure reads
\begin{equation}
  \rho(z)=1-\frac{1}{\pi}
  \arctan\sqrt{\frac{(\lambda-a)(b-z)}{(\lambda-b)(z-a)}} +
  \frac{1}{\pi} \arctan\sqrt{\frac{(\mu-a)(b-z)}{(\mu-b)(z-a)}}
\end{equation}
for $z\in[a,b]$, with $\rho(z)=1$ for $z\in [0,a]\cup[b,\lambda]$.

\subsubsection{The free-energy density}
Expanding the resolvent \eqref{eq:resasymSBS} to order $z^{-2}$, we
get
\begin{multline}
E_\mathrm{SBS}=\frac{1}{8}
    \Big[2\lambda(\lambda+\sqrt{\lambda-a}\sqrt{\lambda-b})
    +2\mu(\mu-\sqrt{\mu-a}\sqrt{\mu-b})
\\ 
-(a+b)
    \left(\sqrt{\mu-a}\sqrt{\mu-b}-\sqrt{\lambda-a}\sqrt{\lambda-b}\right)\Big]
\end{multline}
yielding
\begin{equation}
E_\mathrm{SBS}=\frac{(\lambda-1)(\mu-1)}{x-1}+\frac{1}{2},\qquad x\in[\xc,x_1],
\end{equation}
where the first expression could have as well been obtained from
$E_\mathrm{SBV}$, see \eqref{eq:ESBV} simply by means of the
replacement $\sqrt{\lambda-b}\to - \sqrt{\lambda-b}$. As for the last
expression, it evidently coincides with that for $E_\mathrm{SBV}$, see
the last line of \eqref{eq:ESBV}.  Integration yields the free-energy
density,
\begin{equation}\label{eq:PhiSBS}
  \Phi_\mathrm{SBS}(x)= \frac{1}{2}\log x-(\lambda-1)(\mu-1)\log\frac{x}{x-1},
\end{equation}
whose expression evidently coincide with that in \eqref{eq:PhiSBV} for
the SBV scenario. The integration constant has been fixed to zero by
the sole requirement of continuity of the free-energy density at
$x=x_1$.

In other words, the transition between the SBV and SBS scenarios (when
present, that is only for $\xc<x_1$, or, equivalently, when $\lambda$
and $\mu$ satisfy \eqref{eq:x1gtxc}), does not imply any phase
transition in the model, the free-energy density being the same
function on both sides of $x_1$. The same phenomenon had already been
observed in \cite{CP-15}.  In view of our considerations above, we
denote both $\Phi_\mathrm{SBV}(x)$ and $\Phi_\mathrm{SBS}(x)$ by
$\Phi_\mathrm{I}(x)$.

\subsection{The void-band-void scenario (VBV)}
We recall that in the symmetric case, $\lambda=\mu$, for small values
of $|\log x|$, the relevant scenario was VBV. It is reasonable to
expect the same scenario holds for moderate differences between the
values of $\lambda$ and $\mu$. Let us investigate this possibility,
that is a band $[a,b]$, between two voids, $[0,a]$ and $[b,\lambda]$.

\subsubsection{The resolvent} Inserting the derivative of the potential
\eqref{eq:derivpot} into \eqref{eq:Wsol}, and resorting to
the integration formulae of Appendix \ref{app:integrals}, we obtain
the following expression for the resolvent:
\begin{multline}\label{eq:resasymVBV}
  W(z)=\log\sqrt{x}+\log\frac{\sqrt{a}\sqrt{z-b}+\sqrt{b}\sqrt{z-a}}
  {\sqrt{\lambda-a}\sqrt{z-b}+\sqrt{\lambda-b}\sqrt{z-a}}
\\
  +\log\frac{\sqrt{a}\sqrt{z-b}+\sqrt{b}\sqrt{z-a}}
  {\sqrt{\mu-a}\sqrt{z-b}+\sqrt{\mu-b}\sqrt{z-a}}.
\end{multline}
Requiring the asymptotic behaviour $W(z)\sim 1/z$ as $z\to\infty$,
we obtain the two equations
\begin{equation}\label{eq:eqabasymVBV}
\begin{split}
\frac{\left(\sqrt{\lambda-a}+\sqrt{\lambda-b}\right)
\left(\sqrt{\mu-a}+\sqrt{\mu-b}\right)}
{\big(\sqrt{a}+\sqrt{b}\big)^2}&=\sqrt{x},
  \\ 
\lambda+\mu-\sqrt{\lambda-a}
  \sqrt{\lambda-b}-\sqrt{\mu-a}\sqrt{\mu-b}&=2+2\sqrt{a}\sqrt{b},
\end{split}
\end{equation}
whose solution determines the position of the
end-points $a$ and $b$.

As we shall see below, the resolvent and the end-point equations are
closely related to those of the VBS scenario, discussed in the next
Section. Just as for the SBV/SBS scenarios, the two scenarios VBV and
VBS differ only in a change of sign of $\sqrt{\lambda-b}$, allowing
for a unified treatment, see below.  We also anticipate that, just as
in the SBV/SBS transition, the transition between scenarios VBV and
VBS does not imply a phase transition. Before showing this, we report
here, for completeness, the expression of the equilibrium measure in
the VBV scenario:
\begin{multline}
  \rho(z)=\frac{1}{\pi}
  \arctan\sqrt{\frac{(\lambda-a)(b-z)}{(\lambda-b)(z-a)}}  
+ \frac{1}{\pi} \arctan\sqrt{\frac{(\mu-a)(b-z)}{(\mu-b)(z-a)}}
\\  -\frac{2}{\pi} \arctan\sqrt{\frac{a(b-z)}{b(z-a)}},\qquad z\in[a,b],
\end{multline}
with $\rho(z)=0$ for $z\in [0,a]\cup[b,\lambda]$.

\subsection{The void-band-saturated  scenario (VBS)}
We have seen above that in the symmetric case, for sufficiently small
values of $x$, the relevant scenario is void-band-saturated. Let us
investigate the consequences of such a scenario.  We are thus
considering the scenario with a void $[0,a]$, a band $[a,b]$ and a
saturated interval $[b,\lambda]$. Recall that we assume $\lambda<\mu$.

\subsubsection{The resolvent}
We follow the standard procedure and introduce an auxiliary resolvent
$H(z)$, such that
\begin{equation}
  W(z)=\log\frac{z-b}{z-\lambda}+H(z),
\end{equation}
determined by the auxiliary potential 
\[
U(z)=V'(z)-2\log\frac{b-z}{\lambda-z}, \qquad z\in[a,b].
\]
We obtain
\begin{multline}
  W(z)=\log\sqrt{x} +\log\frac{\sqrt{a}\sqrt{z-b}+\sqrt{b}\sqrt{z-a}}
    {\sqrt{\lambda-a}\sqrt{z-b}-\sqrt{\lambda-b}\sqrt{z-a}}
    \\
    + \log\frac{\sqrt{a}\sqrt{z-b}+\sqrt{b}\sqrt{z-a}}
    {\sqrt{\mu-a}\sqrt{z-b}+\sqrt{\mu-b}\sqrt{z-a}}.
   \end{multline}
Imposing on the resolvent the asymptotic behaviour $W(z)\sim 1/z$, as
$z\to\infty$, yields the two equations
\begin{equation}\label{eq:eqabasymVBS}\begin{split}
  \frac{\sqrt{b}-\sqrt{a}}{\sqrt{b}+\sqrt{a}}
  \frac{\sqrt{\mu-a}+\sqrt{\mu-b}}{\sqrt{\lambda-a}+\sqrt{\lambda-b}}
  &=\sqrt{x},
  \\ \lambda+\mu +\sqrt{\lambda-a}\sqrt{\lambda-b}
  -\sqrt{\mu-a}\sqrt{\mu-b}&=2+2\sqrt{ab},
  \end{split}
\end{equation}
whose solution determines the position of the end-points $a$ and $b$.

Note that the resolvent $W(z)$ and the end-point equations are the
same as in the VBV scenario, see \eqref{eq:resasymVBV} and
\eqref{eq:eqabasymVBV}, except for a change of sign
$\sqrt{\lambda-b}\to-\sqrt{\lambda-b}$.  We shall thus treat these two
scenarios together below.  Before this, let us report, for
completeness, the expression for the equilibrium measure. We have
\begin{multline}
  \rho(z)=\frac{1}{\pi} \arctan\sqrt{\frac{(\mu-a)(b-z)}{(\mu-b)(z-a)}}
  -\frac{1}{\pi}
  \arctan\sqrt{\frac{(\lambda-a)(b-z)}{(\lambda-b)(z-a)}}
\\
  +\frac{2}{\pi} \arctan\sqrt{\frac{b(z-a)}{a(b-z)}},\qquad z\in[a,b],
\end{multline}
with $\rho(z)=0$ for $z\in [0,a]$ and $\rho(z)=1$ for $z\in [b,\lambda]$.

\subsubsection{The end-points}
We now turn to the solution of the end-point equations
\eqref{eq:eqabasymVBV} and \eqref{eq:eqabasymVBS}, for the VBV and VBS
scenarios. As mentioned above, these equations differ just in a sign,
and may be treated together. However, the derivation of their solution
is a bit involved, and we report it in Appendix \ref{app:end-points}.
The resulting expression of the end-points, valid for both the VBV and
the VBS scenarios, may be given in parametric form as follows:
\begin{align}
  a&=A_+(t)+A_-(t)-2\sqrt{A_+(t)A_-(t)}
  \\ b&=A_+(t)+A_-(t)+2\sqrt{A_+(t)A_-(t)}
  \end{align}
where
\begin{align}
  A_+(t)&=\frac{[(2\lambda-1)t+\lambda-\mu][(2\mu-1)t-\lambda+\mu]}
  {2t^2(\lambda+\mu+t)}
  \\
  A_-(t)&=\frac{(t+1)(t-\lambda+\mu)(t+\lambda-\mu)}
  {2t^2(\lambda+\mu+t)}
   \end{align}
and $t$ is determined by the value of $x$, as  the unique root of
\begin{equation}\label{eq:xoft}
  \frac{(1+t)^2(t-\lambda+\mu)(t+\lambda-\mu)}
       {[(2\lambda-1)t+\lambda-\mu][(2\mu-1)t-\lambda+\mu]}=x,
\end{equation}
within the interval $[t_0,\infty)$, with $t_0=\mu-\lambda>0$.

To get some understanding of this result, and to prepare for the
evaluation of the free-energy density, let us discuss some particular
cases.
  \begin{itemize}
\item In the limiting case $x\to 0^+$, the term `$z\log x$' in the
  potential \eqref{eq:potential} becomes the most relevant, and the
  particles tend to fully pack close to the hard wall at
  $z=\lambda$. In this limit the equilibrium measure tends simply to a
  saturated interval $[\lambda-1,\lambda]$, while the band $[a,b]$
  shrinks to zero, with both $a$ and $b$ tending to $\lambda-1$.  This
  translates into $A_+=\lambda-1$, $A_-=0$, which occurs at
  $t_0:=\mu-\lambda$, corresponding, through \eqref{eq:xoft}, to
  $x=0$.
\item It is easily verified that the value of $t\in[t_0,\infty)$ such
  that $x(t)=1$, with $x(t)$ defined by \eqref{eq:xoft} is simply
  $t=\lambda+\mu-2$.
\item The condition $a=0$ corresponds to the left end-point reaching
  the left hard wall, that is to the closure of the left gap $[0,a]$,
  and to its transition from void to saturated (i.e., to the
  transition between VBS and SBS scenarios). It translates here into
  $A_+=A_-$, which occurs at
  \begin{equation}
    \tc=\left(\sqrt{\lambda(\mu-1)} +\sqrt{(\lambda-1)\mu}\right)^2,
  \end{equation}
  which, using \eqref{eq:xoft}, yields
  \begin{equation}
    \xc=\left(\sqrt{\lambda\mu}+\sqrt{(\lambda-1)(\mu-1)}\right)^2,
  \end{equation}
  reproducing \eqref{eq:xcrit2}. Note the nice relation $\tc=\xc-1$.
\item The condition $b=\lambda$ corresponds to the right end-point
  reaching the right hard wall, that is to the closure of the right gap
  $[b,\lambda]$, and to its transition from void to saturated (i.e.,
  to the transition between VBS and VBV scenarios). This occurs when
  $B_+=B_-$, see Appendix \ref{app:end-points}, that is at
\begin{equation}
  t_2=\frac{(\lambda+1)(\mu-\lambda)
    +\sqrt{\lambda(\mu-\lambda)[(\lambda+4)\mu-(\lambda-2)^2]}}
      {2\lambda-1},
\end{equation}
which, using \eqref{eq:xoft}, yields
\begin{multline}\label{eq:x2}
x_2=\frac{1}{2(2\lambda-1)^3(\lambda+\mu-1)}
\Big\{-\lambda^4+2\lambda^3\mu-\lambda^2\mu^2
\\
-10\lambda^3 +10\lambda\mu^2 +12\lambda^2-6\lambda\mu+2\mu^2
-4\mu-4\lambda+2
\\
+\left[(\lambda+4)\mu-(\lambda-2)^2\right]
\sqrt{\lambda(\mu-\lambda)[(\lambda+4)\mu-(\lambda-2)^2]}\Big\}.
\end{multline}
This transition, as already commented, corresponds to a change of sign
of $\sqrt{\lambda-b}$, which does not modify the analytic expression
of the end-points, nor that of the free-energy density; as we shall
see, it describes a change of scenario with no phase transition.
Clearly, increasing the
value of $x$, the VBS scenario will hold until $x$ reaches the lowest
value between $x_2$ and $\xc$. If $x_2<\xc$, the transition occurs at
$x=x_2$, to the VBV scenario. If, instead, $x_2>\xc$, the transition
occurs at $x=\xc$, towards the SBS scenario. In other words, the VBV
scenario may occur only if $x_2<\xc$. In that event it actually holds
for all $x\in[x_2,\xc]$.
\end{itemize}

\subsubsection{The free-energy density}
We now turn to the evaluation of the free-energy density.  We expand
the resolvents \eqref{eq:resasymVBV} and \eqref{eq:resasymVBV} to
order $z^{-2}$, obtaining
\begin{multline}
  E=\frac{1}{8}\left[-2(a+b)\sqrt{ab}+2(\lambda^2+\mu^2)\right.
    \\\left.
    -(a+b+2\mu)
    \sqrt{\mu-a}\sqrt{\mu-b}
           -\nu(a+b+2\lambda)\sqrt{\lambda-a}\sqrt{\lambda-b}\right]
\end{multline}
where $\nu=\pm 1$ for the VBV and VBS scenarios, respectively.
Expressing $E$ in terms of the quantities $A$'s, $B$'s, and $C$'s
defined in Appendix \ref{app:end-points}, and using
\eqref{eq:ABCfinal}, we get
\begin{equation}
  E=\frac{t^2+(8\lambda\mu-3\lambda-3\mu+2)t^2-3(\mu-\lambda)^2t
    +(\mu-\lambda)^2(\lambda+\mu-2)}{8t^2(\mu+\lambda+t)},
\end{equation}
independently of the sign of $\nu$, that is, of the scenario, VBV or
VBS. Now, recalling that $E=x\partial_x\Phi(x)$, we have
\begin{equation}
\Phi(x(t))=\int E\frac{\partial\log x}{\partial t}\rmd t+C,
\end{equation}
where the function $x(t)$ follows from \eqref{eq:xoft}. Evaluating the
integral for VBV and VBS scenarios, we find that 
$\Phi_\mathrm{VBV}=\Phi_\mathrm{VBS}=:\Phi_\mathrm{II}$, 
which reads
\begin{multline}\label{eq:PhiII}
  \Phi_\mathrm{II}(x(t))=\frac{(\lambda+\mu-2)^2}{2}\log t +
  \frac{(\mu-\lambda)^2+2\lambda+2\mu-1}{2}\log(t+1)
  \\
  +\frac{2\lambda-1}{2}\log(t+\lambda-\mu)
  +\frac{2\mu-1}{2}\log(t+\mu-\lambda)
  \\
  -(\lambda+\mu-1)\log(t+\lambda+\mu)
  -\frac{2\lambda^2-2\lambda+1}{2}\log[(2\lambda-1)t+\lambda-\mu]
  \\
  -\frac{2\mu^2-2\mu+1}{2}\log[(2\mu-1)t+\mu-\lambda]+C,
\end{multline}
where 
\begin{multline}
C=\frac{1}{2}\Big[(\lambda-1)^2\log 2(\lambda-1)+(\mu-1)^2\log 2(\mu-1)
\\  
+\lambda^2\log2\lambda+\mu^2\log2\mu\Big]
\end{multline}
is the integration constant, fixed by imposing the
initial condition $\Phi_\mathrm{II}(1)=- \Psi(\lambda-1,\mu-1)$, see
\eqref{eq:phiof1} and \eqref{eq:psiab}, and recalling that
$x(\lambda+\mu-2)=1$. Alternatively one may impose the initial
condition \eqref{eq:phito0a}, recalling that $x\to 0$ corresponds
through \eqref{eq:xoft} to $t\to t_0=\mu-\lambda$. Expression
\eqref{eq:PhiII} is valid for $t\in[\mu-\lambda,\tc]$ (or
  $x\in[0,\xc]$), independently of the value of $t_2$. When $t_2>\tc$
it just does not play any role. When $t_2<\tc$ (or $x_2<\xc$), at $t=t_2$
the transition between scenarios VBS and VBV occurs, with no effects on
the free-energy density.

\subsubsection{The critical point $\xc$ and the phase transition}

Summing up, we have evaluated the free-energy density of the log-gas
in the case $\lambda\not=\mu$, with the following result.  When
$x>\xc$, the free-energy density $\Phi_\mathrm{I}$ is given by
 \eqref{eq:PhiSBV}, or equivalently, by \eqref{eq:PhiSBS}.
When $x<\xc$, the free-energy density $\Phi_\mathrm{II}$ is given by
\eqref{eq:PhiII}.

To study the behaviour of the free-energy density in the vicinity of
the critical point $\xc$, it is convenient to employ the same
parametrization for both $\Phi_\mathrm{I}$ and $\Phi_\mathrm{II}$, and
thus to use parametrization \eqref{eq:xoft}, already used in
$\Phi_\mathrm{II}$, for $\Phi_\mathrm{I}$ as well. We have
\begin{multline}
  \Phi_\mathrm{I}(x(t))=2(\lambda-1)(\mu-1)\log t
  -[2(\lambda-1)(\mu-1)-1]\log(t+1)
  \\
  -\frac{2\lambda\mu-2\lambda-2\mu+1}{2}\log(t+\lambda-\mu)
  -\frac{2\lambda\mu-2\lambda-2\mu+1}{2}\log(t+\mu-\lambda)
  \\
  +(\lambda-1)(\mu-1)\log(t+\lambda+\mu)
  +(\lambda-1)(\mu-1)\log(t-\lambda-\mu+2)
  \\
  -\frac{1}{2}\log[(2\lambda-1)t+\lambda-\mu]
  -\frac{1}{2}\log[(2\mu-1)t+\mu-\lambda].
\end{multline}
It becomes now simply a matter of calculation to verify that both
functions, together with their first and second derivatives, take the
same value at $t=\tc$, while their third derivatives take different
values.  In other words, the model undergoes a third-order phase
transition at $x=\xc$.


\section{Discussion}

We have evaluated the free-energy density $\Phi(x)$ of the log-gas
described by \eqref{eq:tau_log-gas} and \eqref{eq:measure}.  We want
to discuss the obtained results, to compare them with previous works,
and to explain their usefulness in relation to the study of phase
separation phenomena in the five-vertex model.

\subsection{Correspondence between scenarios and regimes}
Two different scaling limits have been considered, namely the case
$\lambda=\mu$, and the case $\lambda\not=\mu$, with quite different
results. In the first case, as $x$ is varied over the range
$x\in[0,\infty)$, there is a perfect correspondence between changes of
  scenarios and phase transitions: we have three possible scenarios,
  each one giving rise to a different expression for the free-energy
  density, and each change of scenario implies a discontinuity in the
  third derivative of the free-energy density, hence a third-order
  phase transition in the model.

In the second case, $\lambda\not=\mu$, the correspondence between
scenarios and phases appears more complicated. Let us first state a few
facts.  When the potential has a hard wall at position $z=\alpha$,
the contact between a band end-point and $\alpha$ implies a change of
scenario, with a transition between void and saturated for the gap in
the vicinity of $\alpha$. Usually this change of scenario comes with
a discontinuity in the third derivative of the free-energy density.

However, if (and only if) together with a hard wall at position
$z=\alpha$, one has a term of the form $|z-\alpha| \log|z-\alpha|$ in
the potential, \emph{with coefficient 1}, the contact between a band
end-point and $\alpha$ still implies a change of scenario, with a
transition between void and saturated for the gap in the vicinity of
$\alpha$, but does not induce any singularity in the free-energy.

Therefore, when $\lambda=\mu$, and the coefficients of the terms
$(\lambda-z)\log(\lambda-z)$ and $z\log z$ are both 2, the
correspondence between changes of scenarios and phase transition is
preserved. Conversely, when $\lambda\not=\mu$, the coefficient of the
term $(\lambda-z)\log(\lambda-z)$, related to the hard wall at
position $\lambda$ becomes equal to 1, and the correspondence between
changes of scenarios and phase transition is broken.  More
specifically, the left hard wall induces a phase transition together
with the change of scenario, while the right does not.

\begin{figure}[t]
\includegraphics{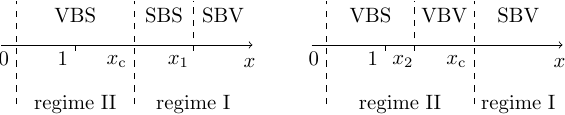}
\caption{Transitions betwen scenarios, and regimes, depending whether
  $x_1,x_2>\xc$ (left) or $x_1,x_2<\xc$ (right). Remarkably, despites
  very different expressions, for given values of $\lambda$ and $\mu$,
  $x_1$ and $x_2$ appears to be both smaller or both larger than
  $\xc$.}\label{fig-phases}
\end{figure}

In relation with these observation, it is worth mentioning the
following fact. It is apparent from their expression, see
\eqref{eq:x1} and \eqref{eq:x2}, that $x_2\not=x_1$. However, and
remarkably, the two condition $x_1>\xc$ and $x_2>\xc$ coincide: they
both select the region \eqref{eq:x1gtxc} in the space of parameters
$\mu>\lambda>1$. This is coherent with the fact that, for given values
of $\lambda$ and $\mu$, the occurrence of the transition from SBV to
SBS (requiring $x_1>\xc$) is not compatible with that from VBS to VBV
(requiring $x_2<\xc$), see Fig.~\ref{fig-phases}.

Indeed, for values of $\lambda$ and $\mu$ satisfying
\eqref{eq:x1gtxc}, as $x$ is decreased within the interval
$[0,\infty)$, one has first the SBV scenario, corresponding to Regime
  I, till the value $x=x_1$ where the scenario changes to SBS, but not
  the regime (no singularity in the free-energy density), and next to
  the value $x=\xc$, where the scenario changes to VBS, with a
  third-order phase transition, to Regime II.

If instead the values of $\lambda$ and $\mu$ do not satisfy
\eqref{eq:x1gtxc}, as $x$ is decreased within the interval
$[0,\infty)$, one has first the SBV scenario, corresponding to Regime
  I, till the value $x=\xc$ where the scenario changes to VBV, with a
  third-order phase transition, to Regime II, and, decreasing $x$
  further, at $x=x_2$ the scenario changes to VBS, still remaining in
  Regime II (no singularity in the free-energy density).

The interpretation provided above for the phases of the log-gas, in
the two different scaling limits, and for their relations with the
various scenarios for the equilibrium measure is complementary to that
provided in \cite{BP-24}, in a five-vertex model picture.

\subsection{Comparison with previous results}
We have evaluated the free-energy density $\Phi(x)$ of the log-gas
described by \eqref{eq:tau_log-gas} and \eqref{eq:measure}.  We want
now to compare the obtained expressions with those worked out in
\cite{BP-24}, using a differential equation approach. To this aim
recall that the free-energy density of the five-vertex model,
$f_2(x)$, see \eqref{eq:deff2}, and that of the log-gas, $\Phi(x)$,
see \eqref{eq:defPhi}, are simply related by \eqref{eq:f2Phi}.

In the language of the five-vertex models, when the scaling limit is
performed for an asymptotically square-shaped domain, $\lambda=\mu$,
depending on the value of $x$, one has three regimes, labelled $I$,
$II$, and $III$, and corresponding to the SBV, the VBV, and the VBS
scenarios, respectively. These three regimes, or scenarios are
separated by two critical points $\xc$, $\xtc=\xc^{-1}$, with $\xc$
given by \eqref{eq:xcrit}. Inserting expressions \eqref{eq:PhisymSBV},
\eqref{eq:PhisymVBV}, and \eqref{eq:PhisymVBS} into \eqref{eq:f2Phi},
one readily recover 
(upon replacement $\lambda \mapsto r+1$) 
the expressions for $f_2^{I}(x)$, $f_2^{II}(x)$,
and $f_2^{III} (x)$, respectively, as provided in \cite{BP-24}, Thm.~1.1.

Similarly, when the scaling limit is performed for an asymptotically
rectangular domain, $\lambda\not=\mu$, depending on the value of $x$,
one has two regimes, labelled $I$, $II$, separated by one critical
point $\xc$ given by \eqref{eq:xcrit2}. Inserting expressions
\eqref{eq:PhiSBV}, or equivalently \eqref{eq:PhiSBS}, and
\eqref{eq:PhiII} into \eqref{eq:f2Phi}, one readily recover 
(upon replacement $\lambda\mapsto p+1$ and $\mu\mapsto q+1$)) the
expressions for $f_2^{I}(x)$ and $f_2^{II}(x)$, respectively, as
provided in \cite{BP-24}, Thm.~1.2.

\subsection{Perspectives}
Experience from the six-vertex model with domain-wall boundary
conditions suggests the possibility of building a representation for
some boundary correlation function of the five-vertex model in terms
of the determinant of a matrix differing from the Hankel matrix
appearing in \eqref{eq:tau_hankel} in just one column. As explained,
e.g., in \cite{CPZj-10}, for suitable forms of this modified column,
it is then possible to relate such a determinantal representation to
the log-gas model associated to \eqref{eq:tau_hankel}. In particular,
the asymptotic behaviour of such a boundary correlation function may
be directly related to the resolvent of the log-gas.

Once the asymptotic behaviour of some boundary correlation function is
explicitly evaluated, the phase separation curves of the model may be
derived by using the `tangent method' \cite{CS-16}. In view of this
the main result of the present paper is indeed the complete derivation
of the free-energy of the five-vertex model within a log-gas
description, and in particular the explicit expression for the
resolvent in the various scenarios.

\section*{Acknowledgments}

We are indebted to Nikolai M.~Bogoliubov, Andrea Maroncelli, 
and Matteo Mucciconi for stimulating discussions. 


\appendix

\section{Asymptotic behaviour of MacMahon's formula}\label{app:psi}

The number of boxed plane partitions is given by MacMahon formula
\eqref{eq:macmahon}. We want here to evaluate
\begin{equation}
  \Psi(a,b):=\lim_{N\to\infty} \frac{1}{N^2}\log \mathrm{PL}(N,\lceil
  aN\rceil,\lceil bN\rceil).
\end{equation}
Evaluating the limit we obtain
\begin{equation}
  \Psi(a,b)=\int_0^1 \left[\ell(a+b+x)-\ell(a+x) -\ell (b+x) +\ell
    (x) \right]\rmd x,
\end{equation}
where we have introduced the notation  
\begin{equation}
\ell(x):=x\log x.
\end{equation}
Standard calculations yields
\begin{multline}
\Psi(a,b)  =\frac{1}{4}
\Big[\ell\left(a^2\right)-\ell\left((a+1)^2\right)
+\ell\left(b^2\right)-
    \ell\left((b+1)^2\right)
 \\
-\ell\left((a+b)^2\right)+\ell\left((a+b+1)^2\right)\Big],
\end{multline}
In particular, we have
\begin{align}
\Psi(0,b)&=\Psi(a,0)=0,
\\ 
\Psi(a,a)&=\frac{1}{2}(1+2a)^2\log(1+2a)
  -(1+a)^2\log(1+a)-a^2\log 4a.
\end{align}
For our purposes, in the main text it is convenient to rewrite the
last expression in terms of $\lambda=a+1$. One has
\begin{multline}\label{eq:Psill}
\Psi\left(\lambda-1,\lambda-1\right) 
=\frac{(2\lambda-1)^2}{2} \log(2\lambda-1)
-\lambda^2\log\lambda
\\
-(\lambda-1)^2\log(\lambda-1)-2(\lambda-1)^2\log 2.
\end{multline}

\section{Useful integrals}\label{app:integrals}

Let us recall a few identities, which turn useful in the
evaluation of the resolvent. The first, obvious one, is simply
\begin{equation}
  \int_a^b\frac{1}{(z-u)\sqrt{(u-a)(b-u)}}\,\rmd u
  =
\frac{\pi}{\sqrt{(z-a)(z-b)}},\qquad z\in\mathbb{C}\backslash[a,b],
\end{equation}
We also have
\begin{multline}
\int_a^b\frac{1}{(z-u)\sqrt{(u-a)(b-u)}}\log\frac{u-c}{u-d}\,\rmd u
\\
=
\begin{cases}
\dfrac{2\pi}{\sqrt{(z-a)(z-b)}}
\log\dfrac{\sqrt{a-c}\sqrt{z-b}+\sqrt{b-c}\sqrt{z-a}}
{\sqrt{a-d}\sqrt{z-b}+\sqrt{b-d}\sqrt{z-a}},\quad
&c,d\leq a,
\\[12pt]
\dfrac{2\pi}{\sqrt{(z-a)(z-b)}}
\log\dfrac{\sqrt{c-a}\sqrt{z-b}+\sqrt{c-b}\sqrt{z-a}}
{\sqrt{d-a}\sqrt{z-b}+\sqrt{d-b}\sqrt{z-a}},\quad
&c,d\geq b,
\end{cases}
\end{multline}
and
\begin{multline}
\int_a^b\frac{1}{(z-u)\sqrt{(u-a)(b-u)}}\log\frac{u-c}{d-u}\,\rmd u
\\
=
\dfrac{2\pi}{\sqrt{(z-a)(z-b)}}
\log\dfrac{\sqrt{a-c}\sqrt{z-b}+\sqrt{b-c}\sqrt{z-a}}
{\sqrt{d-a}\sqrt{z-b}+\sqrt{d-b}\sqrt{z-a}},
\qquad c\leq a,b\leq d,
\end{multline}
which all hold for $z\in\mathbb{C}\backslash[a,b]$.

Finally, we report a useful identity to evaluate the equilibrium
measure from the expression of the resolvent. Let
\begin{equation}
f(z)=\log\left[\sqrt{\alpha(z-a)}+\sqrt{\beta(z-b)}\right],
\end{equation}
then
\begin{equation}
f(x+\rmi 0)-f(x+\rmi 0)=2\rmi \arctan
\sqrt{\frac{\beta(b-x)}{\alpha(x-a)}},\qquad z\in[a,b].
\end{equation}

\section{The solution of the end-point equations (\ref{eq:eqabasymVBV})
  and (\ref{eq:eqabasymVBS})}
\label{app:end-points}

We report here the solution of the end-point equations
\eqref{eq:eqabasymVBV} and \eqref{eq:eqabasymVBS}, for the VBV and VBS
scenarios. We follow an approach proposed in \cite{CP-15}, for a
similar problem.

As mentioned these equations differ just in a sign, and may be treated
together.  To start with, let us rewrite the two sets of end-point
equations in a unified way:
\begin{equation}\label{eq:eqabVBS}
\begin{split}
\frac{\sqrt{b}-\sqrt{a}}{\sqrt{b}+\sqrt{a}}
  \frac{\sqrt{\mu-a}+\sqrt{\mu-b}}{\sqrt{\lambda-a}-\nu\sqrt{\lambda-b}}
  &=\sqrt{x},
  \\ \lambda+\mu -\nu\sqrt{\lambda-a}\sqrt{\lambda-b}
  -\sqrt{\mu-a}\sqrt{\mu-b}&=2+2\sqrt{ab},
\end{split}
\end{equation}
where
\begin{equation}
  \nu=  \begin{cases}
    +1 &\mathrm{for\ scenario\ VBV},\\
    -1 &\mathrm{for\ scenario\ VBS}.
  \end{cases}
\end{equation}
Note that, from the form of the first equation it follows that the
values $\nu=-1$, that is the VBS scenario, corresponds to lower values
of $x$.

Instead of dealing with the end-points $a$ and $b$ as unknowns, we
introduce the quantities
\begin{align}
  A_{\pm}&=\frac{1}{4}\left(\sqrt{b}\pm\sqrt{a}\right)^2,
  \\
  B_{\pm}&=\frac{1}{4}\left(\sqrt{\lambda-a}\pm\nu\sqrt{\lambda-b}\right)^2,
  \\
  C_{\pm}&=\frac{1}{4}\left(\sqrt{\mu-a}\pm\sqrt{\mu-b}\right)^2.
\end{align}
Clearly, the end-points may be expressed in terms of these new
quantities, e.g.,
\begin{equation}
  a=A_{+}+A_{-}-2\sqrt{A_{+}A_{-}},\qquad b=A_{+}+A_{-}+2\sqrt{A_{+}A_{-}},
\end{equation}
and similarly in terms of $B$'s or $C$'s. These quantities
also satisfy some consistency relations, of  multiplicative,
\begin{equation}\label{eq:relABC1}
  A_{+}A_{-}=B_{+}B_{-}=C_{+}C_{-},
\end{equation}
or additive form
\begin{equation}\label{eq:relABC2}
  A_{+}+A_{-}=\lambda-B_{+}-B_{-}=\mu-C_{+}-C_{-}.
\end{equation}
The first equation in \eqref{eq:eqabVBS} reads now
\begin{equation}\label{eq:relABC3}
\frac{A_{-}C_{+}}{A_{+}B_{-}}=x,
\end{equation}
  and the second one can be written in either of the two forms
\begin{equation}\label{eq:relABC4}
2A_{\pm}+B_{\pm}+C_{\pm}=N_{\pm},
\end{equation}
where 
\begin{equation}
N_{+}:=\lambda+\mu-1,\qquad N_{-}:=1.
\end{equation}
Clearly, \eqref{eq:relABC1}--\eqref{eq:relABC4} constitute a system of
six equations in the six unknowns $A_{\pm}$, $B_{\pm}$, and $C_{\pm}$.
Also, the equivalence of two forms of \eqref{eq:relABC4} is easily
checked, just rewriting  \eqref{eq:relABC2} as
\begin{align}\label{eq:relABC5}\begin{split}
  &A_++A_-+B_++B_-=\lambda,\\
  &A_++A_-+C_++C_-=\mu,
    \end{split}
  \end{align}
and summing the two equations.

Let us now turn to the solution of this system of equations.  Dividing
the two equations in \eqref{eq:relABC4} by $\sqrt{B_{\pm}C_{\pm}}$,
respectively, and using the relation $B_{+}/C_{+}=C_{-}/B_{-}$, see
\eqref{eq:relABC1}, comparison of the resulting relations yields
\begin{equation}
\frac{N_{+}-2A_{+}}{\sqrt{B_{+}C_{+}}}=\frac{N_{-}-2A_{-}}{\sqrt{B_{-}C_{-}}}.
\end{equation}
From the ratio of these two relations, and \eqref{eq:relABC1}, we may
now express the $B$'s in terms of the $A$'s and $C$'s, as follows
\begin{equation}\label{eq:CofB}
C_{\pm}=B_{\mp}\frac{w_\pm}{w_\mp},
\end{equation}
where we have used the notation
\begin{equation}\label{eq:Aofz}
w_{\pm} :=N_{\pm}-2A_{\pm}.
\end{equation}
We may now use these relations into \eqref{eq:relABC5}, to eliminate
the $C$'s, and to express the $A$'s in terms of the $w$'s; we get
\begin{align}
  B_{+}+B_{-}&=\frac{w_{+}+w_{-}}{2}-\frac{\mu-\lambda}{2},
    \\
    B_{+}\frac{w_{-}}{w_{+}}+B_{-}\frac{w_{+}}{w_{-}}
    &=\frac{w_{+}+w_{-}}{2}+\frac{\mu-\lambda}{2},
\end{align}
and solving for the $B$'s,
\begin{equation}\label{eq:Bofz}
  B_{\pm}=\left(1\mp\frac{\mu-\lambda}{w_{+}-w_{-}}\right)\frac{w_\pm}{2}.
\end{equation}
Recalling that $B_{+}B_{-}=A_{+}A_{-}$, see \eqref{eq:relABC1}, we have
\begin{equation}
  \left[1-\frac{(\mu-\lambda)^2}{(w_{+}-w_{-})^2}\right]w_{+}w_{-}
  =(N_{+}-w_{+})(N_{-}-w_{-}),
\end{equation}
which turns into a cubic equation for the $w$'s
\begin{equation}
(\mu-\lambda)^2w_{+}w_{-}
=\left(N_{+}w_{-}+N_{-}w_{+}-N_{+}N_{-}\right)(w_{+}-w_{-})^2.
\end{equation}
Note that it contains only terms of degree two and three, thus
describing a rational curve. This means the $w$'s can be expressed as
rational functions of some parameter along the curve.

To obtain this parametrization, we set $w_{+}=(t+1)w_{-}$, and consider $w_{-}$
as a funtion of the parameter $t$; we have
\begin{equation}\label{eq:parametric}
\begin{split}
  w_{+}&=(t+1)w_{-}, 
\\
w_{-}&=\frac{1}{N_{+}+N_{-}(t+1)}
  \left[N_{+}N_{-}+(\mu-\lambda)^2\frac{(t+1)}{t^2}\right].
\end{split}
\end{equation}
Note that we need just the portion of the curve corresponding to the
regime of interest. As we shall see below, this is given by
$t\in[t_0,\infty)$, with $t_0:=\mu-\lambda>0$.

Substituting now \eqref{eq:parametric} into \eqref{eq:CofB},
\eqref{eq:Aofz}, and \eqref{eq:Bofz}, we finally obtain
\begin{equation}\label{eq:ABCfinal}
\begin{split}
    A_{+}&=\frac{[(2\lambda-1)t+\lambda-\mu]
      [(2\mu-1)t-\lambda+\mu]}{2t^2(\lambda+\mu+t)},
    \\
    A_{-}&=\frac{(t+1)(t-\lambda+\mu)(t+\lambda-\mu)}{2t^2(\lambda+\mu+t)},
    \\
    B_{+}&=\frac{(t+1)(t+\lambda-\mu)[(2\lambda-1)t+\lambda-\mu]}
    {2t^2(\lambda+\mu+t)},
    \\
    B_{-}&= \frac{(t-\lambda+\mu)[(2\mu-1)t-\lambda+\mu]}
    {2t^2(\lambda+\mu+t)},
    \\
    C_{+}&=\frac{(t+1)(t-\lambda+\mu)[(2\mu-1)t-\lambda+\mu]}
    {2t^2(\lambda+\mu+t)},
    \\
    C_{-}&=\frac{(t+\lambda-\mu)[(2\lambda-1)t+\lambda-\mu]}
    {2t^2(\lambda+\mu+t)}.
\end{split}
\end{equation}
These expressions for the $A$'s, $B$'s, and $C$'s solve equations
\eqref{eq:relABC1}, \eqref{eq:relABC2}, and \eqref{eq:relABC4}. It can
be checked that they are all positive, as long as
$t\in[t_0,\infty]$.

Note that we have not used equation \eqref{eq:relABC3}, yet. To obtain
the solution for the end-points we need to fix $t\in[t_0,\infty)$.
  This is done indeed by means of \eqref{eq:relABC3}, which using
  \eqref{eq:ABCfinal} yields precisely \eqref{eq:xoft}.  This is a
  quartic equation, with in principle four different roots for any
  given value of $x$. It may be verified that for any given $x>0$,
  there is one and only one real root in $[t_0,\infty)$, and that, as
    $t$ varies over this interval, $x$ increases monotonously,
    spanning the whole interval $[0,\infty)$.

\bibliography{log-gas_bib}

\begin{bibdiv}
\begin{biblist}

\bib{B-82}{book}{
      author={Baxter, R.~J.},
       title={Exactly solved models in statistical mechanics},
   publisher={Academic Press},
     address={San Diego, CA},
        date={1982},
}

\bib{GLT-90}{article}{
      author={Garrod, C.},
      author={Levi, A.~C.},
      author={Touzani, M.},
       title={Mapping of crystal growth onto the 6-vertex model},
        date={1990},
     journal={Solid State Commun.},
      volume={75},
       pages={375\ndash 382},
         doi={10.1016/0038-1098(90)90915-X},
}

\bib{G-90}{article}{
      author={Garrod, C.},
       title={Stochastic models of crystal growth in two dimensions},
        date={1990},
     journal={Phys. Rev. A},
      volume={41},
       pages={4184\ndash 4194},
         doi={10.1103/PhysRevA.41.4184},
}

\bib{GBL-93}{article}{
      author={Gul\'acsi, M.},
      author={van Beijeren, H.},
      author={Levi, A.~C.},
       title={Phase diagram of the five-vertex model},
        date={1993},
     journal={Phys. Rev. E},
      volume={47},
       pages={2473\ndash 2483},
         doi={10.1016/0038-1098(90)90915-X},
}

\bib{NK-94}{article}{
      author={Noh, J.~D.},
      author={Kim, D.},
       title={Interacting domain walls and the five-vertex model},
        date={1994},
     journal={Phys. Rev. E},
      volume={49},
       pages={1943\ndash 1961},
      eprint={cond-mat/9312001},
         doi={10.1103/PhysRevE.49.1943},
}

\bib{HWKK-96}{article}{
      author={Huang, H.~Y.},
      author={Wu, F.~Y.},
      author={Kunz, H.},
      author={Kim, D.},
       title={Interacting dimers on the honeycomb lattice: an exact solution of
  the five-vertex model},
        date={1996},
     journal={Physica A},
      volume={47},
       pages={1\ndash 32},
      eprint={cond-mat/9510161},
         doi={10.1016/S0378-4371(96)00057-X},
}

\bib{MS-13}{article}{
      author={Motegi, K.},
      author={Sakai, K.},
       title={{Vertex models, TASEP and Grothendieck polynomials}},
        date={2013},
     journal={J. Phys. A},
      volume={46},
       pages={355201},
      eprint={1305.3030},
         doi={10.1088/1751-8113/46/35/355201},
}

\bib{MS-14}{article}{
      author={Motegi, K.},
      author={Sakai, K.},
       title={{K}-theoretic boson-fermion correspondence and melting crystals},
        date={2014},
     journal={J. Phys. A},
      volume={47},
       pages={445202},
      eprint={1311.6076},
         doi={10.1088/1751-8113/47/44/445202},
}

\bib{BBBG-19}{article}{
      author={Brubaker, B.},
      author={Buciumas, V.},
      author={Bump, D.},
      author={Gustafsson, H. P.~A.},
       title={Colored five-vertex models and {D}emazure atoms},
        date={2021},
     journal={J. Combin. Theory A},
      volume={178},
       pages={105354},
      eprint={1902.01795},
         doi={10.1016/j.jcta.2020.105354},
}

\bib{M-20}{article}{
      author={Motegi, K.},
       title={{Integrability approach to Feh\'er-N\'emethi-Rim\'anyi-Guo-Sun
  type identities for fac-torial Grothendieck polynomials}},
        date={2020},
     journal={Nucl. Phys. B},
      volume={954},
       pages={114998},
      eprint={1909.02278},
         doi={10.1016/j.nuclphysb.2020.114998},
}

\bib{dGKW-21}{article}{
      author={de~Gier, J.},
      author={Kenyon, R.},
      author={Watson, S.},
       title={Limit shapes for the asymmetric five vertex model},
        date={2021},
     journal={Commun. Math. Phys.},
      volume={385},
       pages={793\ndash 836},
      eprint={1812.11934},
         doi={10.1007/s00220-021-04126-7},
}

\bib{KP-22a}{article}{
      author={Kenyon, R.},
      author={Prause, I.},
       title={The genus-zero five vertex model},
        date={2022},
     journal={Prob. Math. Phys.},
      volume={3},
       pages={707\ndash 729},
      eprint={2101.04195},
         doi={10.2140/pmp.2022.3.707},
}

\bib{KP-22b}{article}{
      author={Kenyon, R.},
      author={Prause, I.},
       title={Gradient variational problem in {$\mathbb{R}^2$}},
        date={2022},
     journal={Duke Math. J.},
      volume={171},
       pages={3003\ndash 3022},
      eprint={2006.01219},
         doi={10.1215/00127094-2022-0036},
}

\bib{KP-24}{article}{
      author={Kenyon, R.},
      author={Prause, I.},
       title={Limit shapes from harmonicity: dominos and the five vertex
  model},
        date={2024},
     journal={J. Phys. A: Math. Theor.},
      volume={57},
       pages={035001},
      eprint={2310.06429},
         doi={10.1088/1751-8121/ad17d7},
}

\bib{KBI-93}{book}{
      author={Korepin, V.~E.},
      author={Bogoliubov, N.~M.},
      author={Izergin, A.~G.},
       title={{Quantum Inverse Scattering Method and Correlation Functions}},
   publisher={Cambridge University Press},
     address={Cambridge},
        date={1993},
}

\bib{B-08}{article}{
      author={Bogoliubov, N.~M.},
       title={Four-vertex model and random tilings},
        date={2008},
     journal={Theor. Math. Phys.},
      volume={155},
       pages={523\ndash 535},
      eprint={0711.0030},
         doi={10.1007/s11232-008-0043-6},
}

\bib{B-10}{article}{
      author={Bogoliubov, N.~M.},
       title={Five-vertex model with fixed boundary conditions},
        date={2010},
     journal={St. Peterburg Math. J.},
      volume={21},
       pages={407\ndash 421},
         doi={10.1090/S1061-0022-10-01100-3},
}

\bib{BM-15}{article}{
      author={Bogolyubov, N.~M.},
      author={Malyshev, C.~L.},
       title={Integrable models and combinatorics},
        date={2015},
     journal={Russian Math. Surveys},
      volume={70},
       pages={789\ndash 856},
         doi={10.1070/RM2015v070n05ABEH004964},
}

\bib{BP-23}{article}{
      author={Bogolyubov, N.~M.},
      author={Pronko, A.~G.},
       title={One-point correlation function of the four-vertex model},
        date={2023},
     journal={J. Math. Sci.},
      volume={275},
       pages={249\ndash 258},
         doi={10.1007/s10958-023-06677-7},
}

\bib{BM-24}{article}{
      author={Bogoliubov, N.~M.},
      author={Malyshev, C.~L.},
       title={Scalar product of the five-vertex model and complete symmetric
  polynomials},
        date={2024},
     journal={J. Math. Sci.},
      volume={284},
       pages={654\ndash 664},
         doi={10.1007/s10958-024-07376-7},
}

\bib{P-16}{article}{
      author={Pronko, A.~G.},
       title={The five-vertex model and enumerations of plane partitions},
        date={2016},
     journal={J. Math. Sci.},
      volume={213},
       pages={756\ndash 768},
         doi={10.1007/s10958-016-2737-x},
}

\bib{BP-21}{article}{
      author={Burenev, I.~N.},
      author={Pronko, A.~G.},
       title={Determinant formulas for the five-vertex model},
        date={2021},
     journal={J. Phys. A: Math. Theor.},
      volume={54},
       pages={055008},
      eprint={2011.01972},
         doi={10.1088/1751-8121/abd785},
}

\bib{M-24}{misc}{
      author={Maroncelli, A.},
       title={Limit shape phenomena in statistical mechanics},
        date={2024},
        note={PhD thesis, University of Florence (unpublished)},
}

\bib{K-82}{article}{
      author={Korepin, V.~E.},
       title={Calculations of norms of {B}ethe wave functions},
        date={1982},
     journal={Commun. Math. Phys.},
      volume={86},
       pages={391\ndash 418},
         doi={10.1007/BF01212176},
}

\bib{I-87}{article}{
      author={Izergin, A.~G.},
       title={Partition function of the six-vertex model in the finite volume},
        date={1987},
     journal={Sov. Phys. Dokl.},
      volume={32},
       pages={878\ndash 879},
}

\bib{BPZ-02}{article}{
      author={Bogoliubov, N.~M.},
      author={Pronko, A.~G.},
      author={Zvonarev, M.~B.},
       title={Boundary correlation functions of the six-vertex model},
        date={2002},
     journal={J. Phys. A: Math. Gen.},
      volume={35},
       pages={5525\ndash 5541},
      eprint={math-ph/0203025},
         doi={10.1088/0305-4470/35/27/301},
}

\bib{CS-16}{article}{
      author={Colomo, F.},
      author={Sportiello, A.},
       title={Arctic curves of the six-vertex model on generic domains: The
  tangent method},
        date={2016},
     journal={J. Stat. Phys.},
      volume={164},
       pages={1488–1523},
      eprint={1605.01388},
         doi={10.1007/s10955-016-1590-0},
}

\bib{BP-24}{article}{
      author={Burenev, I.~N.},
      author={Pronko, A.~G.},
       title={Thermodynamics of the five-vertex model with scalar-product
  boundary conditions},
        date={2024},
     journal={Commun. Math. Phys.},
      volume={405},
       pages={148},
      eprint={2312.17565},
         doi={10.1007/s00220-024-05021-7},
}

\bib{KP-16}{article}{
      author={Kitaev, A.~V.},
      author={Pronko, A.~G.},
       title={Emptiness formation probability of the six-vertex model and the
  sixth {P}ainlev{\'e} equation},
        date={2016},
     journal={Commun. Math. Phys.},
      volume={345},
       pages={305\ndash 354},
      eprint={1505.00032},
         doi={10.1007/s00220-016-2636-5},
}

\bib{JM-81}{article}{
      author={Jimbo, M.},
      author={Miwa, T.},
       title={Monodromy preserving deformation of linear ordinary differential
  equations with rational coefficients. {II}},
        date={1981},
     journal={Physica D},
      volume={2},
       pages={407\ndash 448},
         doi={10.1016/0167-2789(81)90021-X},
}

\bib{O-87}{article}{
      author={Okamoto, K.},
       title={{Studies on the Painlev\'e equations. I: sixth Painlev\'e
  equation $P_{\mathrm{VI}}$}},
        date={1987},
     journal={Ann. Mat. Pura Appl.},
      volume={146},
       pages={337\ndash 381},
         doi={10.1007/BF01762370},
}

\bib{Zj-00}{article}{
      author={Zinn-Justin, P.},
       title={Six-vertex model with domain wall boundary conditions and
  one-matrix model},
        date={2000},
     journal={Phys. Rev. E},
      volume={62},
       pages={3411\ndash 3418},
      eprint={math-ph/0005008},
         doi={10.1103/PhysRevE.62.3411},
}

\bib{CPZj-10}{article}{
      author={Colomo, F.},
      author={Pronko, A.~G.},
      author={Zinn-Justin, P.},
       title={The arctic curve of the domain-wall six-vertex model in its
  anti-ferroelectric regime},
        date={2010},
     journal={J. Stat. Mech. Theory Exp.},
      volume={2010},
       pages={L03002},
      eprint={1001.2189},
         doi={10.1088/1742-5468/2010/03/L03002},
}

\bib{F-10}{book}{
      author={Forrester, P.~J.},
       title={Log-gases and random matrices},
      series={London Mathematical Society Monographs},
   publisher={Princeton University Press},
     address={Princeton, NJ},
        date={2010},
      volume={34},
}

\bib{BCMP-23}{article}{
      author={Burenev, I.~N.},
      author={Colomo, F.},
      author={Maroncelli, A.},
      author={Pronko, A.~G.},
       title={Arctic curves of the four-vertex model},
        date={2023},
     journal={J. Phys. A: Math. Theor.},
      volume={55},
       pages={465202},
      eprint={2307.03076},
         doi={10.1088/1751-8121/ad02ce},
}

\bib{BKMM-07}{book}{
      author={Baik, J.},
      author={Kriecherbauer, T.},
      author={McLaughlin, K. T.-R.},
      author={Miller, P.~D.},
       title={{Discrete orthogonal polinomials: Asymptotics and applications}},
      series={Ann. of Math. Stud.},
   publisher={Princeton University Press},
     address={Princeton, NJ},
        date={2007},
      volume={164},
}

\bib{DK-93}{article}{
      author={Douglas, M.~R.},
      author={Kazakov, V.~A.},
       title={{Large $N$ phase transition in continuum QCD$_2$}},
        date={1993},
     journal={Phys. Lett. B},
      volume={319},
       pages={219\ndash 230},
      eprint={hep-th/9305047},
         doi={10.1016/0370-2693(93)90806-S},
}

\bib{DS-00}{article}{
      author={Dragnev, P.~D.},
      author={Saff., E.~B.},
       title={A problem in potential theory and zero asymptotics of
  {K}rawtchouk polynomials},
        date={2000},
     journal={J. Approx. Theory},
      volume={102},
       pages={120\ndash 140},
         doi={10.1006/jath.1999.3366},
}

\bib{CP-15}{article}{
      author={Colomo, F.},
      author={Pronko, A.~G.},
       title={Thermodynamics of the six-vertex model in an {L}-shaped domain},
        date={2015},
     journal={Comm. Math. Phys.},
      volume={339},
       pages={699\ndash 728},
      eprint={1501.03135},
         doi={10.1007/s00220-015-2406-9},
}

\end{biblist}
\end{bibdiv}

\end{document}